\documentclass[journal, twocolumn, 10pt]{IEEEtran}
\usepackage{amssymb,dcolumn,delarray}
\usepackage{hyperref}
\usepackage{epstopdf}
\usepackage{pdfpages}
\usepackage{float}
\usepackage{graphicx}
\usepackage{cite}
\usepackage{tabularx}
\usepackage{amsmath}
\usepackage{mathtools}
\usepackage{dsfont}
\usepackage[binary-units = true, range-phrase=--, per-mode=symbol]{siunitx}
\usepackage{enumitem}
\usepackage{booktabs}

\usepackage{color}
\usepackage{colortbl}
\newcolumntype{L}[1]{>{\raggedright\arraybackslash}p{#1}}
\newcolumntype{C}[1]{>{\centering\arraybackslash}p{#1}}
\newcolumntype{R}[1]{>{\raggedleft\arraybackslash}p{#1}}

\usepackage{array}
\newcolumntype{?}{!{\vrule width 1pt}}
\usepackage{makecell}

\DeclareSIUnit{\decibelm}{dBm}
\DeclareSIUnit{\Joule}{Joule}

\ifCLASSINFOpdf
\else
\fi
\ifCLASSOPTIONcompsoc
 \usepackage[caption=false,font=normalsize,labelfont=sf,textfont=sf]{subfig}
\else
 \usepackage[caption=false,font=footnotesize]{subfig}
\fi
\begin{document}
%
\title{Understanding the Performance of Bluetooth Mesh: Reliability, Delay and Scalability Analysis}
%
%
%

\author{Ra{\'u}l~Rond{\'o}n,~\IEEEmembership{Member,~IEEE,}
        Aamir~Mahmood,~\IEEEmembership{Senior~Member,~IEEE,}
        Simone~Grimaldi,
        and~Mikael~Gidlund,~\IEEEmembership{Senior~Member,~IEEE}
\thanks{R. Rond\'{o}n, A. Mahmood, S. Grimaldi, and M. Gidlund are with the Department of Information Systems and Technology, Mid Sweden University, Sundsvall 851 70, Sweden (e-mail: raul.rondon@miun.se; aamir.mahmood@
miun.se; simone.grimaldi@miun.se;
mikael.gidlund@miun.se).}
}

\maketitle

\begin{abstract}
This article evaluates the quality-of-service performance and scalability of the recently released Bluetooth Mesh protocol and provides general guidelines on its use and configuration. Through extensive simulations, we analyzed the impact of the configuration of all the different protocol's parameters on the end-to-end reliability, delay, and scalability. In particular, we focused on the structure of the packet broadcast process, which takes place in time intervals known as \textit{Advertising Events} and \textit{Scanning Events}. Results indicate a high degree of interdependence among all the different timing parameters involved in both the scanning and the advertising processes and show that the correct operation of the protocol greatly depends on the compatibility between their configurations. We also demonstrated that introducing randomization in these timing parameters, as well as varying the duration of the \textit{Advertising Events}, reduces the drawbacks of the flooding propagation mechanism implemented by the protocol. Using data collected from a real office environment, we also studied the behavior of the protocol in the presence of WLAN interference. It was shown that Bluetooth Mesh is vulnerable to external interference, even when implementing the standardized limitation of using only 3 out of the 40 Bluetooth Low Energy frequency channels. We observed that the achievable average delay is relatively low, of around 250~ms for over 10 hops under the worst simulated network conditions. However, results proved that scalability is especially challenging for Bluetooth Mesh since it is prone to broadcast storm, hindering the communication reliability for denser deployments.

\end{abstract}


%
\IEEEpeerreviewmaketitle

\section{Introduction}
%
%
%
%
\IEEEPARstart{O}{ver} the past few years, Bluetooth Low Energy (BLE) has become one of the most widely-adopted technologies for a great variety of applications such as wearable devices, health care, home automation and Internet-of-Things (IoT)~\cite{Raza, Zhang, 7000963}. BLE is a wireless technology developed by the Bluetooth Special Interest Group (SIG)~\cite{SIG} and was officially introduced in 2010 with the version 4.0 of the Bluetooth Core Specification. It provides an ultra-low power consumption performance, suited for applications involving power-limited devices, and has even proved to be capable of meeting the stringent quality-of-service (QoS) requirements of single-hop real-time industrial applications~\cite{7794553, Rondon2017}. Even though the protocol is mature and has established a strong position in the market,  limited network size and lack of standardized multi-hop models prevent BLE from occupying a key position in applications outside the short-range umbrella.

For over a decade now, Internet-of-Things (IoT) has been one of the main topics explored in both industry and academia. As a result, IoT devices and applications have both increased and diversified exponentially. A significant increase in the number of devices in networks, as well as in the coverage, is a common trend in most recent applications. In these cases, Ad Hoc mesh networks are preferred over the single failure point nature of the centralized star topology of BLE~\cite{7968828}. Consequently, several research efforts have been made to mitigate this issue by exploiting the capability of a device to perform simultaneously as a master of a network and as a slave of others. Many proprietary mesh schemes based on BLE have been proposed, e.g., see~\cite{Patti, 7889350, 8328554}. Even though these solutions are potentially viable, they lack the interoperability, which characterizes standardized network technologies, which results in compatibility issues that limit their penetration in the market.

Released in July 2017, Bluetooth Mesh is the latest technology developed by the Bluetooth SIG. It uses the existing BLE protocol stack to allow for many-to-many device communication over BLE radio. In contrast to the operational independence between Bluetooth Basic Rate/Enhanced Data Rate and BLE, the Bluetooth mesh functionality is completely based and dependent on the BLE protocol stack. Therefore, it is more accurate to view Bluetooth mesh as a networking technology than a wireless communication technology~\cite{MeshWeb}. Bluetooth Mesh enables the creation of mesh networks consisting of thousands of devices, therefore greatly increasing the coverage range. It follows a connection-less flooding mechanism, in which a packet is broadcast by each device receiving it. The simplicity in its structure, the low cost of implementation, and interoperability with the widely available BLE make Bluetooth Mesh an attractive technology for most mid-range IoT applications. 

Being a new technology, not many studies focusing on the evaluation of Bluetooth Mesh have been presented thus far. An overview and experimental evaluation of the protocol is described in~\cite{MeshFirstOverview}. In this study, reliability and delay performances are analyzed based on the number of hops between the sender and the destination. However, the effects of the different parameters and network design choices involved are not extensively addressed.  The drawbacks of using a flooding model were addressed and diverse approaches to select the devices in the network to take on the relay role were proposed in~\cite{MeshOnRelaySelection}. More recently, a model of the current consumption, lifetime and energy cost per delivered bit was presented in~\cite{s19051238}.

In this article, we analyze the impact of choosing different configurations of the Bluetooth Mesh protocol parameters on the resulting end-to-end (e2e) reliability and delay performance. We mainly focus on the configuration of the \textit{Advertising Events} and \textit{Scanning Events} timing, including the \textit{ScanInterval} and $T_{\mathrm{interPDU}}$, as well as the result of introducing randomization in the time parameters. We also study the behavior of the protocol for increasing density of devices deployed over a given area, as well as the effect of self and external interference in the final performance. All of the involved parameters are described in detail in Section \ref{Section3}. The key contribution of this work can be summarized as follows. 

\begin{enumerate}
    \item To the best of our knowledge, this study is the first to analyze in detail the interdependence among all the different Bluetooth Mesh protocol's parameters. We showed that the correct operation of the protocol depends on the compatibility between the configuration of both the scanning and the advertising processes. 
    \item We identified the key elements of the protocol that should be considered when designing a scalable Bluetooth Mesh network. In particular, we evaluated the effect of the flooding propagation mechanism, as well as the configuration of the protocol's parameters in the final QoS performance.
    \item We showed that Bluetooth Mesh is vulnerable to external interference, even when implementing the limited scheme with only 3 out of the 40 BLE frequency channels. This shows that the use of different sets of frequency channels, other than the standard BLE advertising channels, represents a possible tool for performance optimization.
    \item Overall, we provide general guidelines for the use and configuration of the protocol. Being a new technology, not much is yet known about Bluetooth Mesh, and the available description offers a very flexible definition of the many parameters. This study contributes towards a better understanding of the protocol, serving as a basis for future studies.
\end{enumerate}

The remainder of this article is organized as follows. Section II provides an overview of the Bluetooth Mesh protocol stack and structure. Section III describes the concepts related to the BLE ADV process and explains how these are used in Bluetooth Mesh. Section IV presents the evaluation of the performance and limitations of Bluetooth Mesh, in terms of different configurations of the network and protocol parameters, as well as interference conditions. Finally, Section V concludes this article.

\section{Background} \label{MeshIntro}
   
    Bluetooth mesh networking follows a publish/subscribe messaging system model. When \textit{Publishing}, devices may send messages to specific unicast or multicast addresses. Devices can also be configured to receive messages sent to particular addresses by other devices, also known as \textit{Subscribing}. Messages published to a given address will be received by all other devices subscribed to that particular address. As an example, a switch node can publish an ON/OFF message to an address named "Kitchen", then all lights subscribed to that address will receive the message and react accordingly. Each node possesses a subscriber list with its unicast address, which is given to every single node in the network, as well as the multicast group address that the device is subscribed to.
    \subsection{Protocol Stack}
    Bluetooth Mesh is a networking protocol that uses the BLE protocol stack as its functional core. As defined in the Mesh Profile specification~\cite{MeshProfile}, the  Bluetooth Mesh protocol presents a layered architecture shown in Table~\ref{tab:StackMesh}. Each layer has its own functions and responsibilities, and it provides certain services to the layer above it. 
\definecolor{Gray}{gray}{0.9}
\definecolor{Gray1}{gray}{0.7}
\bgroup
\centering
\def\arraystretch{1.4}
\begin{table}[t]
\centering
	\caption{Bluetooth mesh protocol stack.}
	\centering
		\begin{tabular}{?C{0.29\linewidth}?p{0.60\linewidth}?}
			\Xhline{2\arrayrulewidth}
		\textbf{Protocol Stack} &  \textbf{Function} \\
		\Xhline{2\arrayrulewidth}
		\multicolumn{2}{c}{}\\[-1.4em]
		\Xhline{2\arrayrulewidth}
		{Model Layer} & Defines standardized application models such as lightning and sensor\\
		\Xhline{2\arrayrulewidth}
		\multicolumn{2}{c}{}\\[-1.4em]
		\Xhline{2\arrayrulewidth}
		{Foundation  Model  Layer} & Defines models to configure and manage a mesh network \\
		\Xhline{2\arrayrulewidth}
		\multicolumn{2}{c}{}\\[-1.4em]
		\Xhline{2\arrayrulewidth}
		{Access layer} &  Defines format of application data, controls data encryption/decryption, maintains network context and application keys\\
		\Xhline{2\arrayrulewidth}
		\multicolumn{2}{c}{}\\[-1.4em]
		\Xhline{2\arrayrulewidth}
		{Upper Transport Layer} & Handles encryption, decryption and authentication of application data\\
		\Xhline{2\arrayrulewidth}
		\multicolumn{2}{c}{}\\[-1.4em]
		\Xhline{2\arrayrulewidth}
		{Lower Transport Layer} & Handles message segmentation and reassembly of transport PDUs \\
		\Xhline{2\arrayrulewidth}
		\multicolumn{2}{c}{}\\[-1.4em]
		\Xhline{2\arrayrulewidth}
		{Network Layer} & Defines addressing of transport messages, rules to relay/accept/reject messages, network message format, and encryption/authentication\\
		\Xhline{2\arrayrulewidth}
		\multicolumn{2}{c}{}\\[-1.4em]
		\Xhline{2\arrayrulewidth}
		{Bearer Layer} &  Defines how to exchange messages among nodes using ADV and GATT bearers, where ADV carries most of the traffic in the network\\
		\Xhline{2\arrayrulewidth}
		\multicolumn{2}{c}{}\\[-1.4em]
		\Xhline{2\arrayrulewidth}
		\rowcolor{Gray}
		\textbf{BLE Core Specification} &  Bluetooth low energy stack (see Section~\ref{MeshIntro}) \\
		\Xhline{2\arrayrulewidth}
		\end{tabular}
	\label{tab:StackMesh}
\end{table}
\egroup

    \subsection{Nodes Features and Roles}
    According to the Bluetooth mesh specification, a node may possess different features that allow each node to play special roles in the network. The defined features are the following.
   \begin{itemize}
      \item \textbf{Relay:}
      Relay Nodes receive and retransmit Bluetooth mesh messages, using the advertising (ADV) bearer. This is one of the most important features since the relay process is what makes Bluetooth mesh multi-hop communication possible.
      \item \textbf{Friend:}
     Friendship is one of the new concepts introduced with the Bluetooth mesh standard and was designed to assist devices with particularly energy-limited characteristics. A Friend Node (FN) is in charge of storing and forwarding, upon request, the messages which are destined to the associated Low Power Nodes.
      \item \textbf{Low Power Node (LPN):}
      Even though Bluetooth mesh is based on BLE, it is more energy-demanding due to the nature of mesh networking. However, a Low Power Node (LPN) can operate in a Bluetooth mesh with the aid of a FN. The LPN can function as a mostly-off device and request the messages stored by its associated FN during its off period.
     \item \textbf{Proxy:}
      A Proxy Node is capable of receiving messages over a given bearer, both ADV or GATT, and retransmit them over the other ADV or GATT bearers.
    \end{itemize}
    \subsection{Network Topology}
\begin{figure}[t]
      \centering
      \includegraphics[width = 1\linewidth]{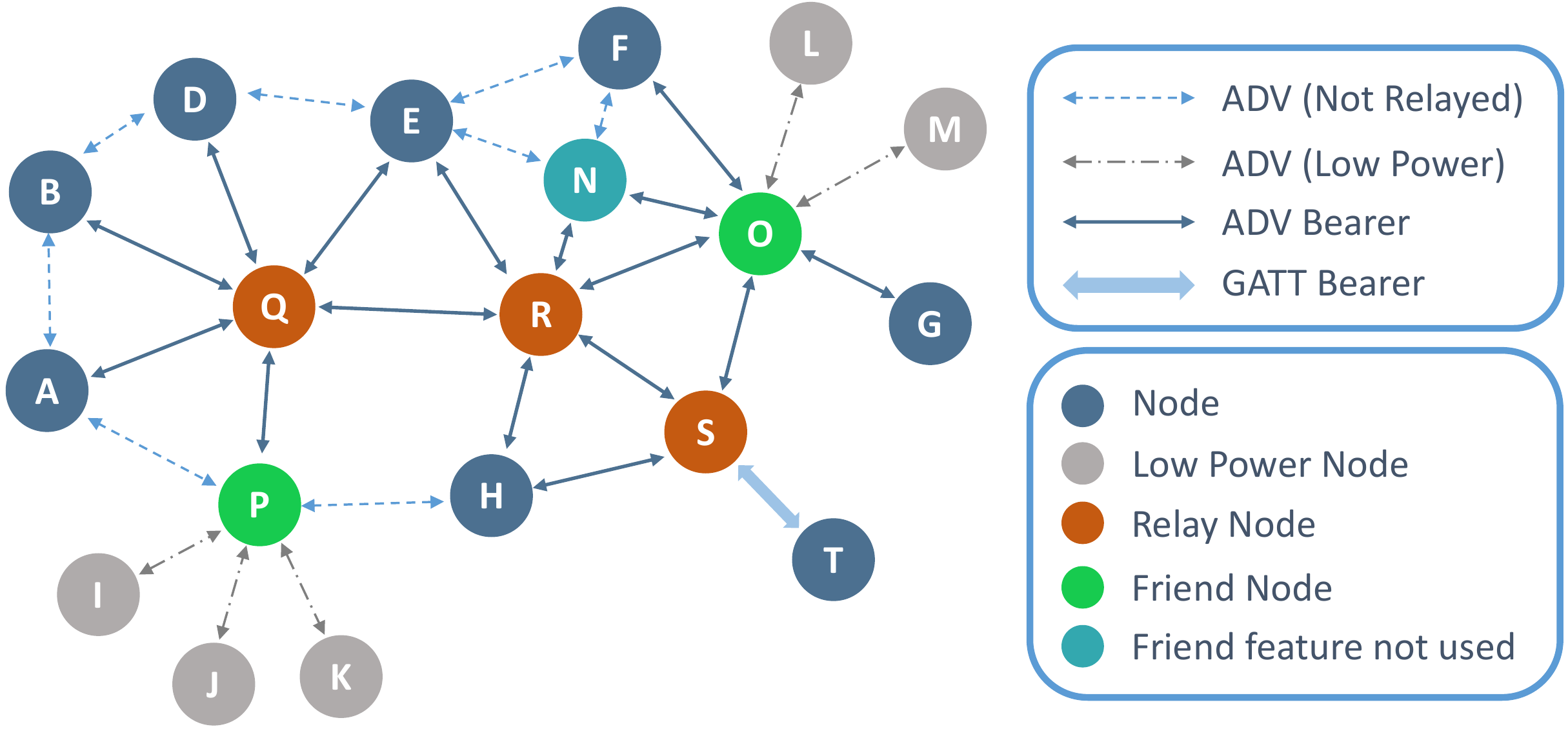}
      \caption{Bluetooth Mesh network topology.}
      \label{MeshTopo}
      \end{figure}

    Figure \ref{MeshTopo} shows an example of a Bluetooth mesh network that is already fully provisioned, i.e., all devices have been accepted into the network and assigned the correspondent addresses and network keys. In this network, the nodes capable of relaying messages are Q, R, and S. Nodes I, J, K, L, and M are LPNs, and thus they are required to establish a friendship with a neighbor node that has the friendship feature enabled. The friendship enabled nodes are P, N, and O, with O being the friend of L and M, and P being the friend of I, J, and K. Node N, although having the friendship feature enabled, is not making use of it. All messages destined to nodes I, J, and K will be received and stored by node P, while node O will do the same for nodes L and M. Nodes A, B, C, D, E, F, G, H are receiving only nodes, whereas node T is connected to the network only through the GATT bearer.  


\section{BLE Broadcast-Oriented Communication}
\label{Section3}
Bluetooth mesh is designed to be used without the need to establish connections among devices in the network. To achieve this connection-less approach, all data transmissions in a Bluetooth Mesh network are broadcast under the advertising/scanning scheme of BLE. 

BLE defines three frequency channels---Channels 37, 38, and 39---reserved for all non-connected state communications. During the advertising process, packets are broadcast sequentially over the three ADV frequency channels within the span of an \textit{Advertising Event}. Each packet transmission is spaced by a time interval of up to 10~ms, hereafter defined as $T_{\textrm{interPDU}}$. These packets can be received by all devices in radio range scanning on a matching ADV channel. The scanning process is conducted within \textit{Scanning Events} that periodically start after a time interval known as \textit{ScanInterval}. At the beginning of a \textit{Scanning Event}, the scanning device starts listening on the next ADV channel from the last channel used. The duration in which the Link Layer scans on one ADV channel is defined by the \textit{ScanWindow} parameter. If the \textit{ScanWindow} is equal to the \textit{ScanInterval}, the device is continuously scanning. The maximum value possible for the \textit{ScanInterval} is 10.24 s.
\begin{figure}[t]
      \centering
      \includegraphics[scale = 0.4]{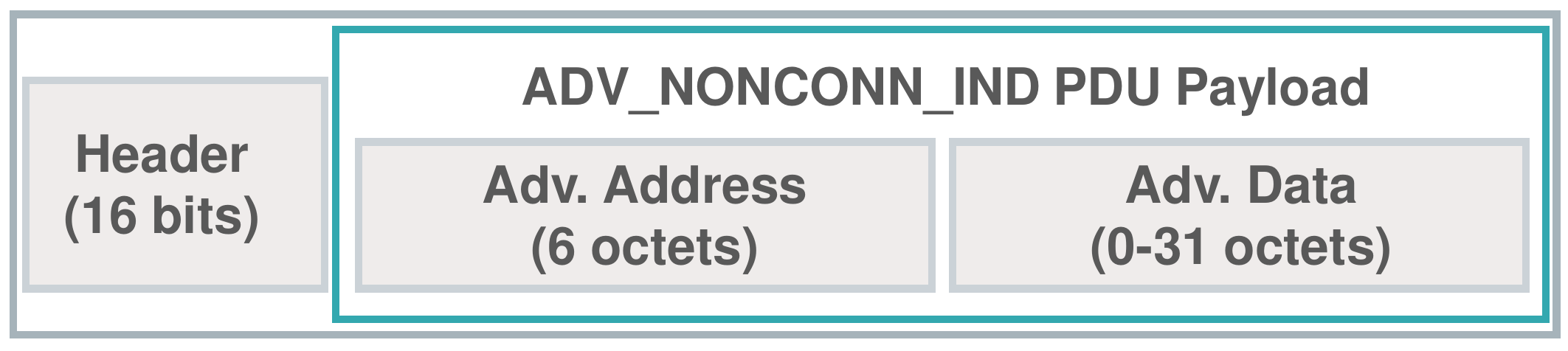}
      \caption{Bluetooth Mesh PDU format.}
      \label{fig:NONCONNPDU}
\end{figure}
\begin{figure*}[h]
      \centering
      \includegraphics[width=0.8\linewidth]{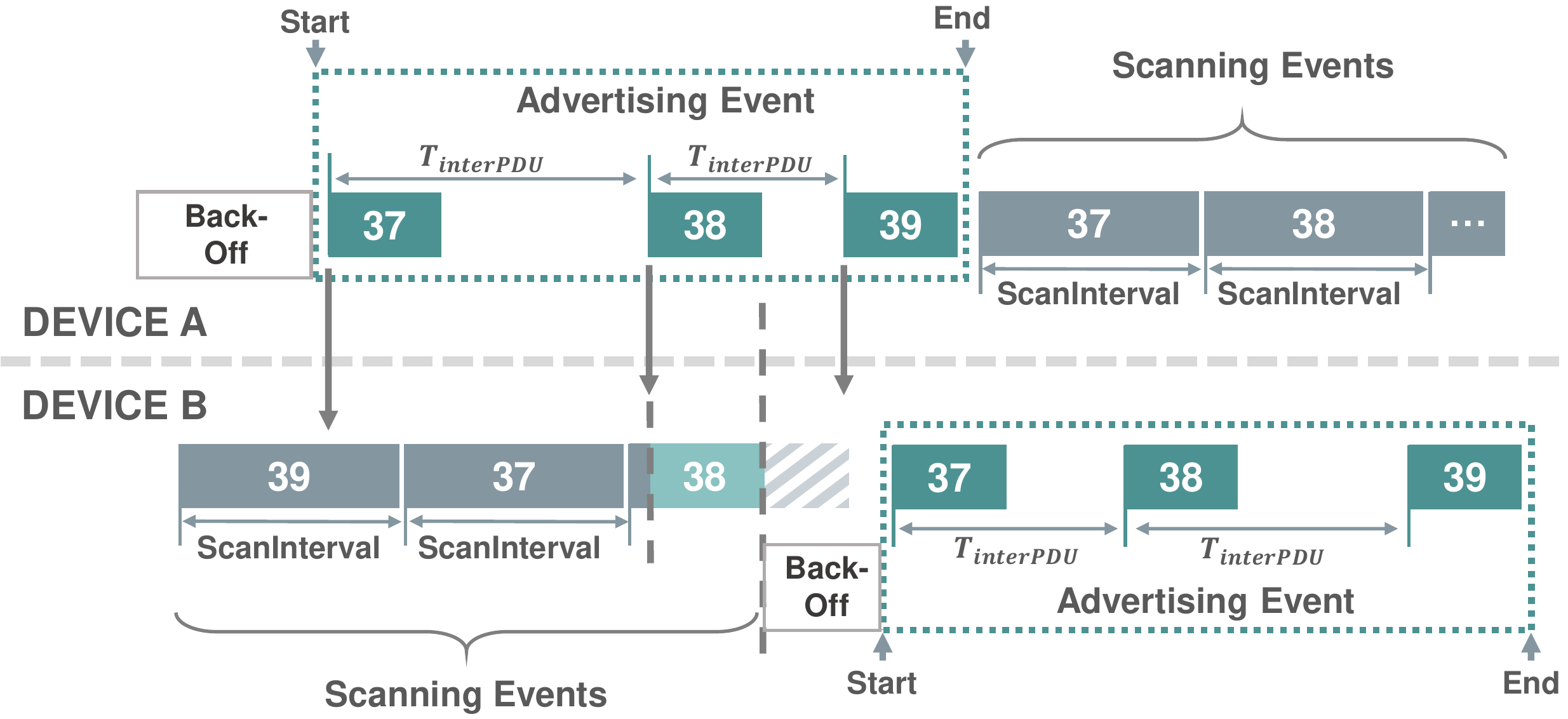}
      \caption{Example of communication flow between two relay devices in a Bluetooth Mesh network.}
      \label{fig:meshDiagram}
\end{figure*}
The Bluetooth Core Specification \cite{SIG} defines several types of \textit{Advertising Events}, as well as the corresponding packet formats, designed to fit specific operational scenarios. Bluetooth Mesh uses \textit{non-connectable undirected event types}, which allows a scanner to receive information from the advertiser, but does not allow them to respond to other requests from scanners or initiators. As shown in Fig. \ref{fig:NONCONNPDU}, the \textit{ADV\_NONCONN\_IND PDU} packet format has a maximum size of 39 octets.

\subsection{Bluetooth Mesh Approach}
    The Bluetooth Mesh transmission model is based on \textit{managed flooding}. In this scheme, all nodes receive all messages from nodes that are in direct radio range and, in case these are relay nodes, the message will be broadcast again to reach nodes that are more distant to the source node. The managed flooding implements a \textit{time-to-live (TTL)} limitation to control the maximum number of hops, over which a message is relayed. Furthermore, all nodes implement a message cache, which contains all of the recently seen messages. When a message is found in the cache, it is considered to have been processed before and it is immediately discarded. 
 
    The use of managed flooding allows messages to reach their destination via multiple paths through the network, potentially improving reliability performance. However, flooding can also increase collision probability. To counteract this, Bluetooth Mesh offers flexibility in the system configuration. It is possible to include random back-off periods before the start of an \textit{Advertising Event}, as well as to randomize the separation $T_{\mathrm{interPDU}}$ between the three transmissions within the \textit{Advertising Event}.
 
    When randomness is introduced in the configuration of the aforementioned protocol parameters, the alignment of the states of different devices cannot be easily predicted. An example of a communication flow is depicted in Fig. \ref{fig:meshDiagram}. In this example, devices A and B are both relays, in radio range, and device A broadcasts a protocol data unit (PDU) which should be received by device B. After a random back-off, the \textit{Advertising Event} of device A starts while device B is scanning an ADV channel. The first transmission from A is not received by B because the latter is not listening to the channel that A is using. Both devices coincide on channel 38 during the second transmission from A, thus device B receives the packet. Assuming that the PDU should be relayed, device B will broadcast it during the \textit{Advertising Event} that starts after a new random back-off. Device A, on the other hand, proceeds to transmit the PDU on the third ADV channel and then enters scanning state until it has new data to transmit.

\section{Bluetooth Mesh Performance Evaluation}
 In order to evaluate the QoS performance and limitations of Bluetooth Mesh, we analyzed the impact of choosing different configurations for the network and protocol parameters on the e2e scalability, reliability and delay performance. We base our analysis on the e2e packet loss rate, the e2e delay, and the  number of packet errors per unit time or Packet Error Rate (PER) per link. In particular, we focused on studying:
\begin{enumerate}[label=(\Alph*)]
	\item The effect of using randomization of the time separation between consecutive transmissions within an \textit{Advertising Event} ($T_{\mathrm{interPDU}}$), as well as the use of PDU repetitions, on the e2e packet loss rate and average delay.
	\item The interdependence among the configuration of scanning events and ADV events. More specifically, we analyzed the interaction among the duration of the \textit{Advertising Events}, the $T_{\mathrm{interPDU}}$, and the $scanInterval$.
	\item The scalability potential of a Bluetooth Mesh network, as well as the limitations that derive from the adoption of the controlled-flooding broadcast approach.
	\item The coexistence of a Bluetooth Mesh network in heterogeneous scenarios. In particular, we evaluated the performance of the protocol under the presence of WLAN interference.
\end{enumerate}
 \begin{figure}[h]
      \centering
      \includegraphics[scale=0.3]{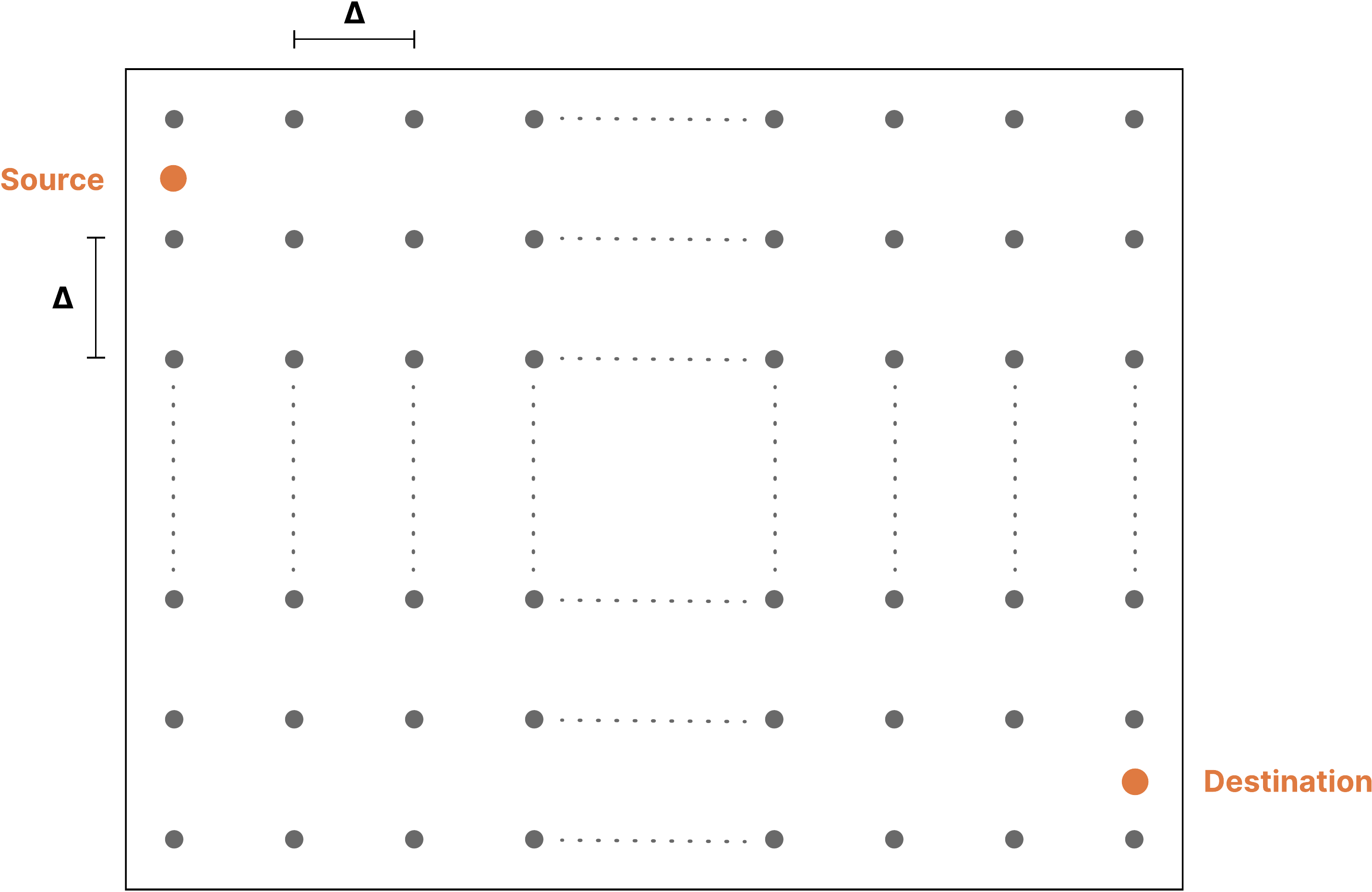}
      \caption{Structure of the grid topology used.}
      \label{fig:meshGrid}
\end{figure}
 \begin{figure*}[h]
    \centering
    \begin{minipage}{0.49\textwidth}
        \includegraphics[width=0.9\textwidth]{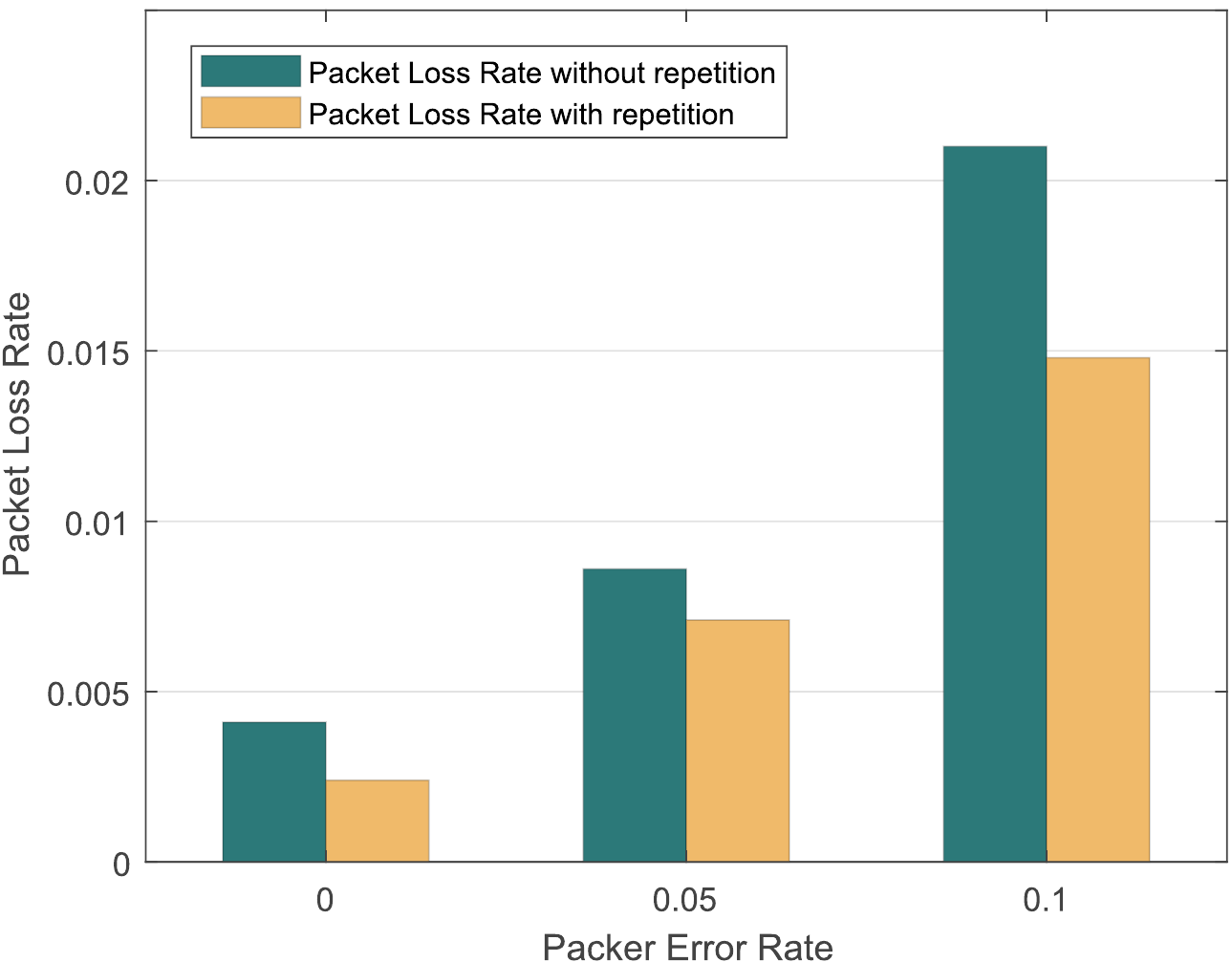} 
        \caption{Effect of using a single packet repetition on the end-to-end packet loss rate.}
        \label{MeshR1}
    \end{minipage}\hfill
    \begin{minipage}{0.49\textwidth}
        \includegraphics[width=0.89\textwidth]{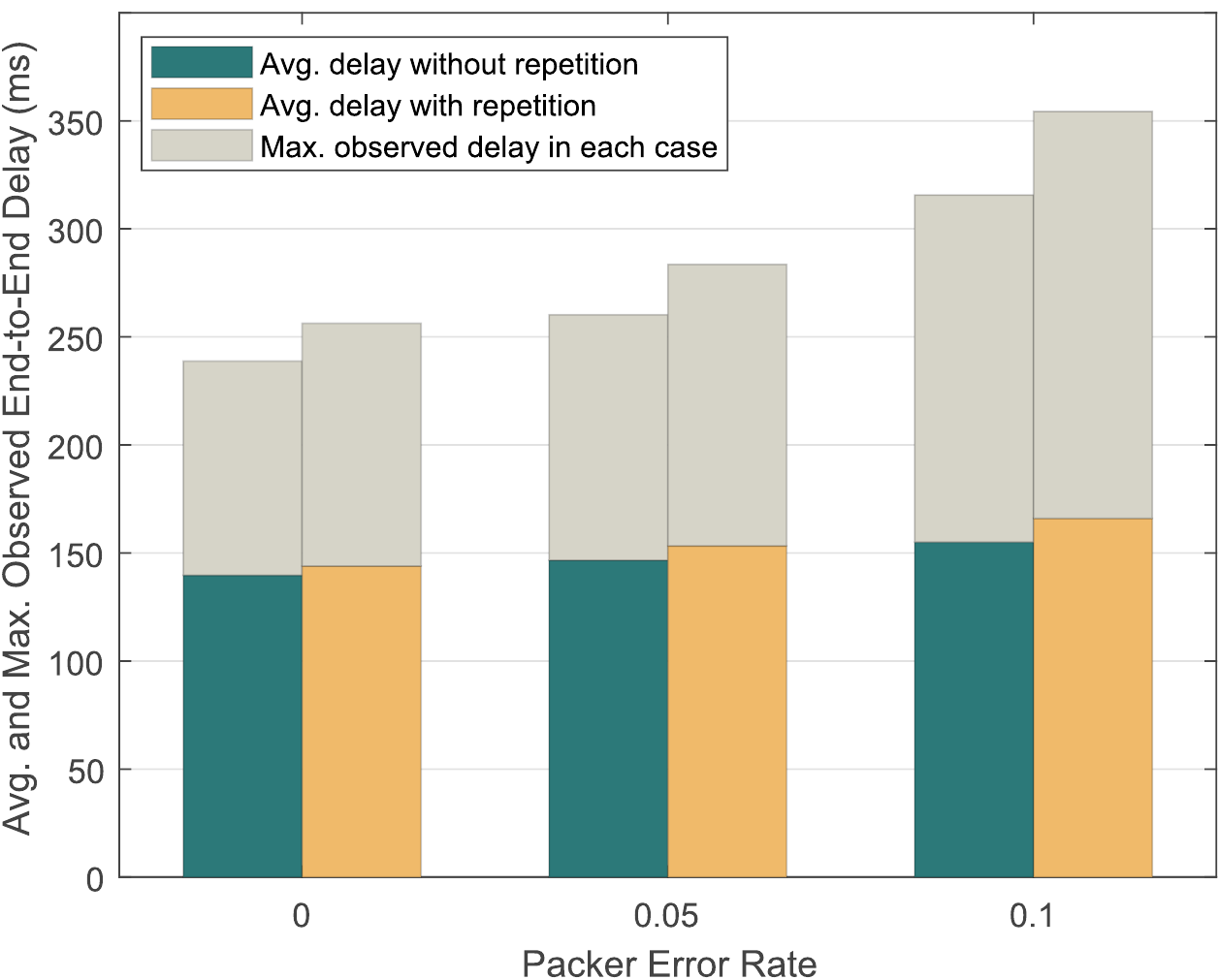} 
        \caption{Effect of using a single packet repetition on the average and maximum delay.}
        \label{MeshR2}
    \end{minipage}
\end{figure*}
\begin{table}[h]
\setlength\extrarowheight{6pt}
\caption{Simulation Parameters for test cases A,B,C, and D}
\label{SimulationTable}
\resizebox{\linewidth}{!}{%
\begin{tabular}{|c|c|c|c|c|}
\hline
\textbf{Parameter}   & \textbf{A}   & \textbf{B}                                                          & \textbf{C}         & \textbf{D} \\ \hline
$\Delta$ (m)         & \multicolumn{2}{c|}{8}                                                             & 8.8, 8, 7.2, 6.6   & -          \\ \hline
Number of devices    & \multicolumn{2}{c|}{68}                                                            & 52, 68, 86, 106    & 28         \\ \hline
R (m)                & \multicolumn{4}{c|}{9}                                                                                               \\ \hline
PDU size             & \multicolumn{4}{c|}{39 octets}                                                                                       \\ \hline
TTL                  & \multicolumn{4}{c|}{127}                                                                                             \\ \hline
PER                  & 0, 0.05, 0.1 & 0                                                                   & 0, 0.05, 0.1 0.15  & -          \\ \hline
Repetition time (ms) & \multicolumn{4}{c|}{30}                                                                                              \\ \hline
Random back-off (ms) & \multicolumn{4}{c|}{0-20}                                                                                            \\ \hline
ScanInterval (ms)    & 20           & 10-200                                                              & \multicolumn{2}{c|}{20}         \\ \hline
$T_{\mathrm{interPDU}}$ (ms)  & 3-5          & \begin{tabular}[c]{@{}c@{}}1-2, 2-4, 4-6, \\ 6-8, 8-10\end{tabular} & \multicolumn{2}{c|}{0.1-1, 3-5} \\ \hline
\end{tabular}%
}
\end{table}
For the sake of generality, we based our analysis of (A), (B), and (C) on a grid topology shown in Fig. \ref{fig:meshGrid}, in which all devices act as relay nodes. The radio range $R$ is considered to be 9 m and the maximum PDU length of 39 octets was used. Since we are studying the e2e performance, the inclusion of scanner-only devices would not result in significant findings. In addition to this, when using flooding as a message propagation method, increasing the number of devices acting as a relay can lead to higher network congestion and packet loss. This is a particularly important aspect to consider since the Bluetooth Mesh documentation advises to use unacknowledged communication. 

For the study of the performance under interference (D), we used WLAN interference data collected from a real office environment. We designed a Bluetooth Mesh network in this office space and simulated its performance under the recorded channel conditions.

The results presented in the following sections were obtained through extensive simulations, of 10 thousand runs for each experiment, using a MATLAB simulation tool we developed for this purpose. The parameters involved in the different simulations are summarized in table \ref{SimulationTable}.

\subsection{Use of Timing Randomness and PDU Repetitions}\label{MeshS2}
As previously mentioned, flooding propagation can increase the probability of collision for denser deployments. The two main tools provided by the Bluetooth Mesh protocol to improve the reliability performance are the introduction of randomness in the timing parameters and transmitting replicas of the PDU of interest. Great freedom is given by the protocol for these functional design choices and not many guidelines are available. It is therefore important to study the cost and results of using them. 

When a relay device broadcasts a PDU, all devices within radio range should be able to hear this transmission. If more than one device receives this PDU correctly, the PDU is not in the message cache, and the TTL is larger than 1, all these devices will then attempt to further relay it. Therefore, introducing a random back-off before the start of a new \textit{Advertising Event} is necessary, as mentioned by Baert et al. in \cite{MeshFirstOverview}. This random back-off range is not specified in the protocol documentation and is left open for each implementation. In all the experiments hereby presented we used a random back-off of up to 20~ms. 

As seen in Fig. \ref{fig:meshDiagram}, the three PDU transmissions on the different ADV channels are spaced by the $T_{\mathrm{interPDU}}$ parameter. In order to meet the timing limitations of the \textit{non-connectable undirected event types}, the $T_{\mathrm{interPDU}}$ can take on a maximum value of 10~ms. To further reduce the collision probability, the protocol documentation advises randomizing  $T_{\mathrm{interPDU}}$, allowing for more desynchronization among transmissions. We have considered this parameter to be randomized in different value ranges and is explained in the following subsection. 

The use of the random back-off and the randomization of the $T_{\mathrm{interPDU}}$ do not present significant drawbacks that hinder the network behavior. On the other hand, transmitting replicas of the PDUs can potentially increase the transmission delay. If each relay node, which is reached by the PDU of interest, is set to transmit one or more replicas, these devices will be occupied handling the current PDU for considerably longer periods of time. Considering that each replica will require an \textit{Advertising Event}, the total handling time can increase significantly, resulting in higher network congestion. For this reason, we considered transmitting only a single replica of the PDU by the initial source device to be a viable configuration. 

To test the effect of transmitting one replica by the initial source device, we performed a simulation in which we observed the behavior of the protocol when sending the single replica and when sending the PDU without any repetition. For this case, we used the grid topology shown in Fig. \ref{fig:meshGrid} with $\Delta = 8$~m, resulting in a total of 68 devices. We also considered that the replica is sent 30~ms after the initial broadcast by the source device. The $T_{\mathrm{interPDU}}$ range was set from 3~ms to 5~ms and, as general values, we assumed three different values of Packet Error Rate (PER) of 0, 0.05 and 0.1, respectively.

The resulting packet loss rate is presented in Fig. \ref{MeshR1}. It is shown that by sending one replica of the original PDU, as expected, the packet loss rate is significantly reduced by up to 25\%. This is particularly visible for higher PER values. On the other hand, it can be seen in Fig. \ref{MeshR2} that the average transmission delay is slightly higher when using the replica transmission. In terms of the average delay only, it can be said that this increase is an acceptable trade-off for the improvement in reliability. The maximum observed delay, however, is considerably higher when sending the PDU replica. This is expected since if the packet delivered to the destination devices is the replica of the original PDU, it will arrive at least 30~ms later than if it was the original PDU itself. The maximum delay in Bluetooth Mesh is not deterministic unless it is strictly limited by the TTL, and therefore this larger increase in the maximum observed delay is not significant enough to discourage the use of the single replica. Moreover, by using a shorter time interval between the original PDU and the replica transmissions, this result could be further improved.

\subsection{Configuration of Scanning Events and Advertising Events} \label{suucsecmeshtim}

Randomization of the $T_{\mathrm{interPDU}}$ parameter separating the three PDU transmissions performed within the \textit{Advertising Event} is recommended by the Bluetooth Mesh protocol specification. This is done with the goal of reducing the probability of two devices simultaneously attempting to broadcast in the same ADV channel. The basic limitation of this parameter is that it should be shorter than 10~ms. The use of shorter $T_{\mathrm{interPDU}}$ values translates to a shorter \textit{Advertising Event} duration, and consequently in a potentially shorter transmission delay. Furthermore, shorter \textit{Advertising Events} can also reduce network congestion since devices will finish handling a single PDU faster. However, using short $T_{\mathrm{interPDU}}$ also implies it is more likely the three different transmissions will encounter the same, or closely similar, radio channel conditions. This is an important aspect to consider, especially in heterogeneous deployments in which the channels are prone to instantaneous interference spikes.

Broadcast PDUs are received by the neighbor devices, which are scanning the corresponding ADV channel. Thus, when choosing the \textit{Advertising Event} configuration it is important to understand the relationship between the \textit{Advertising Event} and the \textit{Scanning Event} timings. If the scanning devices change the active ADV channel too frequently in comparison to the structure of the \textit{Advertising Event} of the broadcaster, the probability that none of the three PDU transmissions within the \textit{Advertising Event} finds a neighbor scanning on a matching channel, increases. In contrast, if the \textit{ScanInterval} is too large for the current \textit{Advertising Event}, it is more likely that one of the broadcast PDUs will be received. This, however, also makes the system more susceptible to failures due to unfavorable channel conditions since a device remains in the same channel for longer time periods. Moreover, since Bluetooth Mesh uses only three radio channels for all packet transmissions, channel usage is a highly important element to consider.

In order to analyze the interdependence of \textit{Scanning Event} and \textit{Advertising Event} timing parameters, we extensively simulated different combinations of these parameters using the same topology used in section \ref{MeshS2}, with the transmission of a single replica of the original PDU, but with PER equal to 0. In this case, we used \textit{ScanInterval} values in the range from 10~ms to 200~ms, and five different random $T_{\mathrm{interPDU}}$ values within ranges: $1-2$~ms, $2-4$~ms, $4-6$~ms, $6-8$~ms, and $8-10$~ms. 
\begin{figure}[t]
      \centering
      \includegraphics[width=1\linewidth]{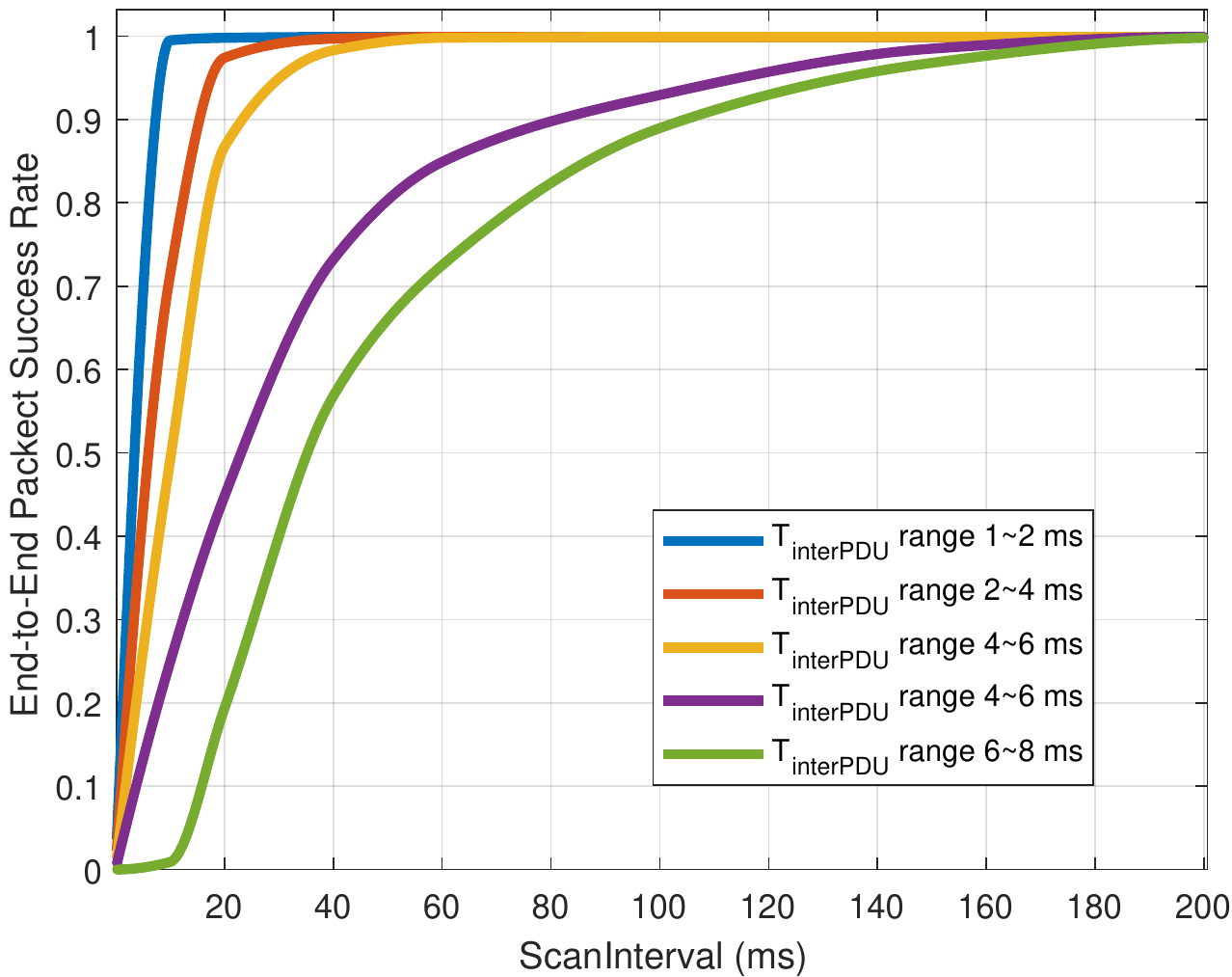}
      \caption{Packet success rate for different configurations and combinations of $ScanInterval$ and $T_{\mathrm{interPDU}}$ values.}
     \label{fig:meshR3}
\end{figure}

The obtained results, shown in Fig. \ref{fig:meshR3}, indicate that when using the two larger $T_{\mathrm{interPDU}}$ ranges of $6\sim8$~ms and $8\sim10$~ms, the e2e packet success rate, or the complement of the packet loss rate, observed for the smaller \textit{ScanInterval} ranges considered is remarkably low. In particular, for  a $T_{\mathrm{interPDU}}$ value from 8 to 10~ms, the use of a 20~ms \textit{ScanInterval} produces a packet success rate below 20\%. This confirms that for larger inter PDU separation, and therefore longer \textit{Advertising Event} duration, much larger \textit{ScanInterval} at the receiving side are required. For the particular values used in this experiment, a \textit{ScanInterval} larger than 200~ms is needed to avoid the system failure due to these timing issues when using a larger $T_{\mathrm{interPDU}}$. It can also be noted that for smaller $T_{\mathrm{interPDU}}$ values, the system reaches a close to 100\% packet success rate for much smaller \textit{ScanInterval} values. For the case of $T_{\mathrm{interPDU}}$ within 1 and 2~ms, for example, the system successfully works with the smallest \textit{ScanInterval} value of 10~ms. Different application requirements can determine the allowed range for $T_{\mathrm{interPDU}}$ and a suitable \textit{Scanning Event} configuration is mandatory to allow the correct network behavior.

\subsection{Scalability of Bluetooth Mesh Networks}
In recent mid-to-long range IoT applications, the number of devices in a network can reach large numbers as massive device deployments are becoming more popular. In this regard, the use of the controlled-flooding broadcast scheme can represent a challenge for the scalability of a Bluetooth Mesh network. Increasing the number of relay devices provides alternate possible paths for a PDU to be successfully transmitted through the mesh network. In other words, the availability of these routes potentially contributes to improving the overall reliability performance. However, in networks using flooding mechanisms, one of the main issues that can be encountered is known as broadcast storm \cite{8359778}. It represents the increase of collision probability, network congestion, and packet loss rate, due to the number of broadcast transmissions.
 \begin{figure*}[h]
    \centering
    \begin{minipage}{0.49\textwidth}
        \includegraphics[width=0.9\textwidth]{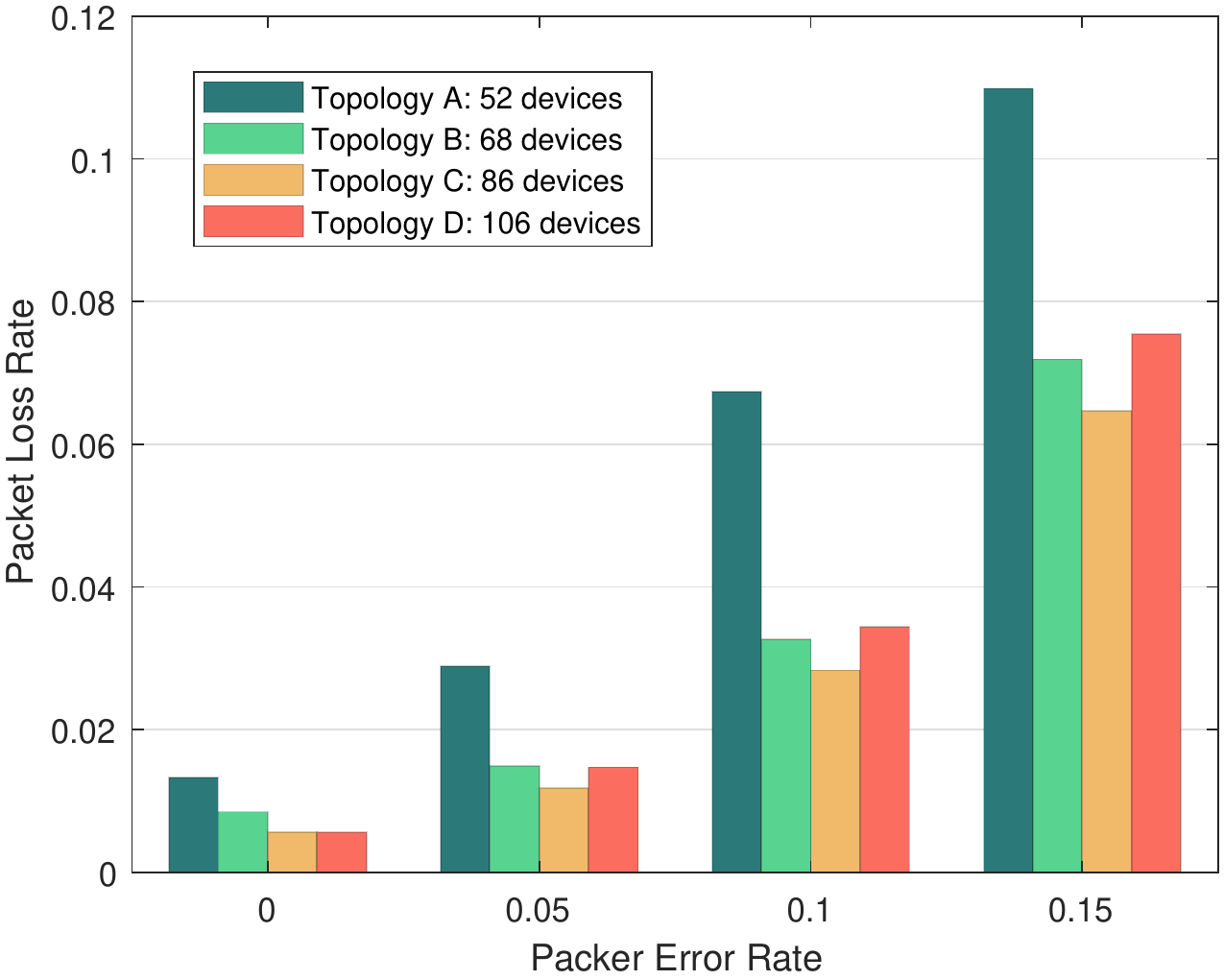} 
        \caption{End-to-end packet loss rate for increasing number of devices.}
        \label{MeshR41}
    \end{minipage}\hfill
    \begin{minipage}{0.49\textwidth}
        \includegraphics[width=0.9\textwidth]{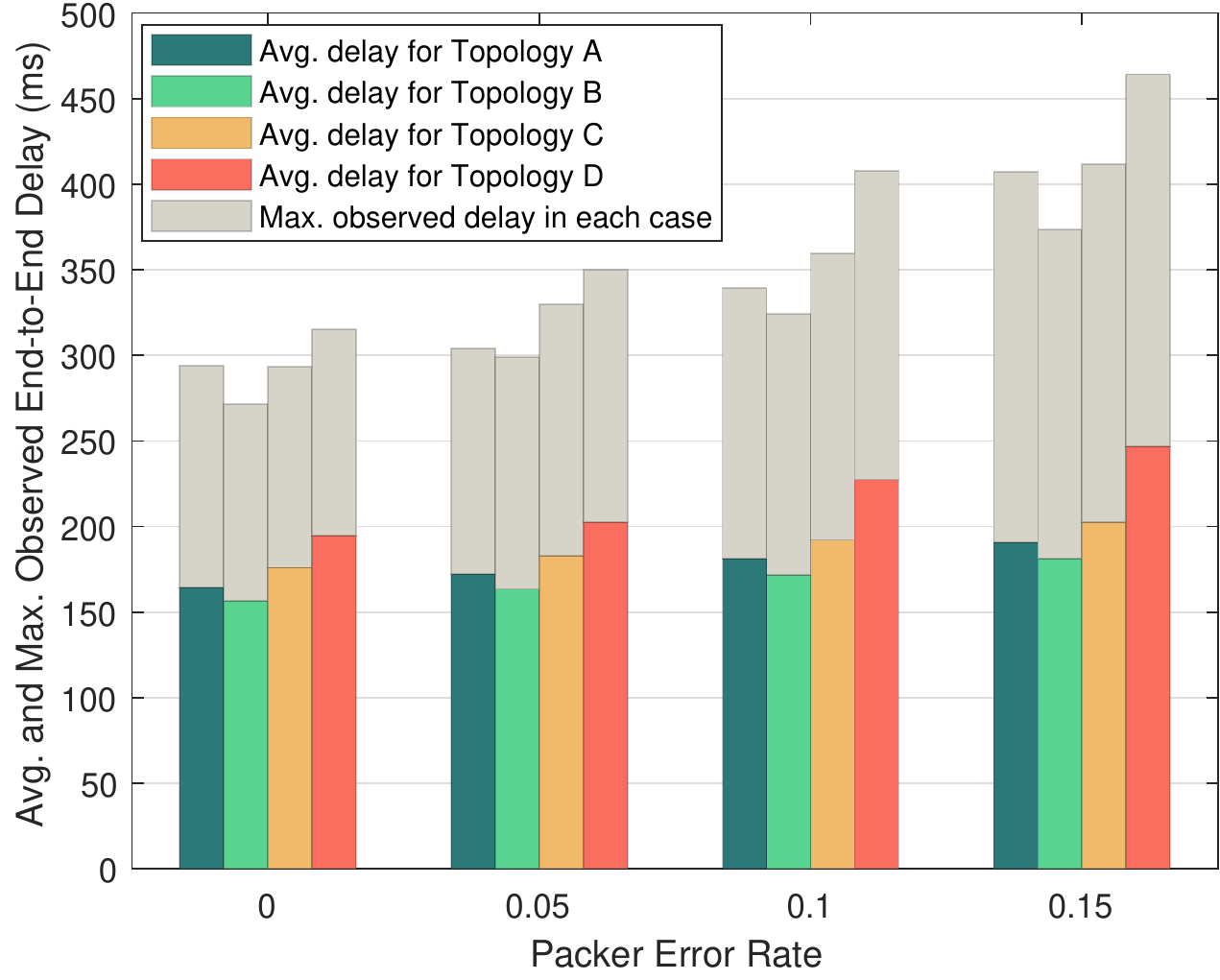} 
        \caption{End-to-end avg. and max. delay for increasing number of devices.}
        \label{MeshR42}
    \end{minipage}
\end{figure*}

In this section, we study the behavior of a Bluetooth Mesh network for increasing numbers of devices in terms of e2e packet loss rate and delay. For this purpose, we simulated the transmission of a single PDU in a mesh network with a grid topology as depicted in Fig. \ref{fig:meshGrid}, with one repetition sent by the source device and no other data traffic in the network. We used four different network topologies by varying the separation $\Delta$ between neighboring devices. In particular, we considered $\Delta$ to be 8.8~m, 8~m, 7.2~m, and 6.6~m, which resulted in Topology A with 52, Topology B with 68, Topology C with 86, and Topology D with 106 devices, respectively. We also considered PER values of 0, 0.05, 0.1, and 0.15.

\subsubsection{PDU Transmission Without Side-Traffic}
The e2e packet loss rate resulting from the different scenarios is shown in Fig. \ref{MeshR41}. It can be seen that Topology A produces the highest packet loss rate in all cases, which suggests the density of devices was not sufficient to provide enough routes for the packet to be delivered successfully. Topology B, with a larger number of devices, shows a vast improvement in comparison to Topology A. This trend is also followed by Topology C, which has an even denser deployment. However, with Topology D the packet loss rates starts increasing again. This shows that after a certain limit, having too many relay devices in the network hinders the data transmission process. As network congestion increases, the positive effect of having more paths for the PDU to travel through becomes overshadowed.

In terms of the delay performance,  Fig. \ref{MeshR42} shows there is an improvement from Topology A to Topology B, but thereafter both the average and the maximum observed delay continue to increase. In the case of Topology A, the average delay is slightly higher than in Topology B due to the fewer and longer paths the PDUs can go through. In contrast, in the case of Topology C, the presence of many devices within radio range allows the packet to hop several times, potentially more than necessary. This is more evident for the case of Topology D, in which the average and maximum delay increase considerably. 

Bluetooth Mesh devices are considered to be limited in computational resources, and thus the use of routing mechanisms or neighbor awareness is not included in the protocol specification. However, the use of such mechanisms could help in counteracting the issues observed for denser networks.  
\vspace{1.5mm}
\subsubsection{PDU Transmission With Side-Traffic}

We repeated the experiment with the same configuration as previously explained but in the presence of more data traffic in the network. We set 10\% of the devices to actively broadcast their own data packets as side traffic in addition to the PDU sent from the source device to the destination device, as shown in Fig. \ref{fig:meshGrid}. Moreover, we simulated two different scenarios to show the effect of the $T_{\mathrm{interPDU}}$ value in the presence of more network traffic. In particular, we used a small $T_{\mathrm{interPDU}}$ in the range from 0.1 to 1~ms, and a larger $T_{\mathrm{interPDU}}$ from 3~ms to 5~ms.
 \begin{figure*}[h]
    \centering
    \begin{minipage}{0.49\textwidth}
        \includegraphics[width=0.9\textwidth]{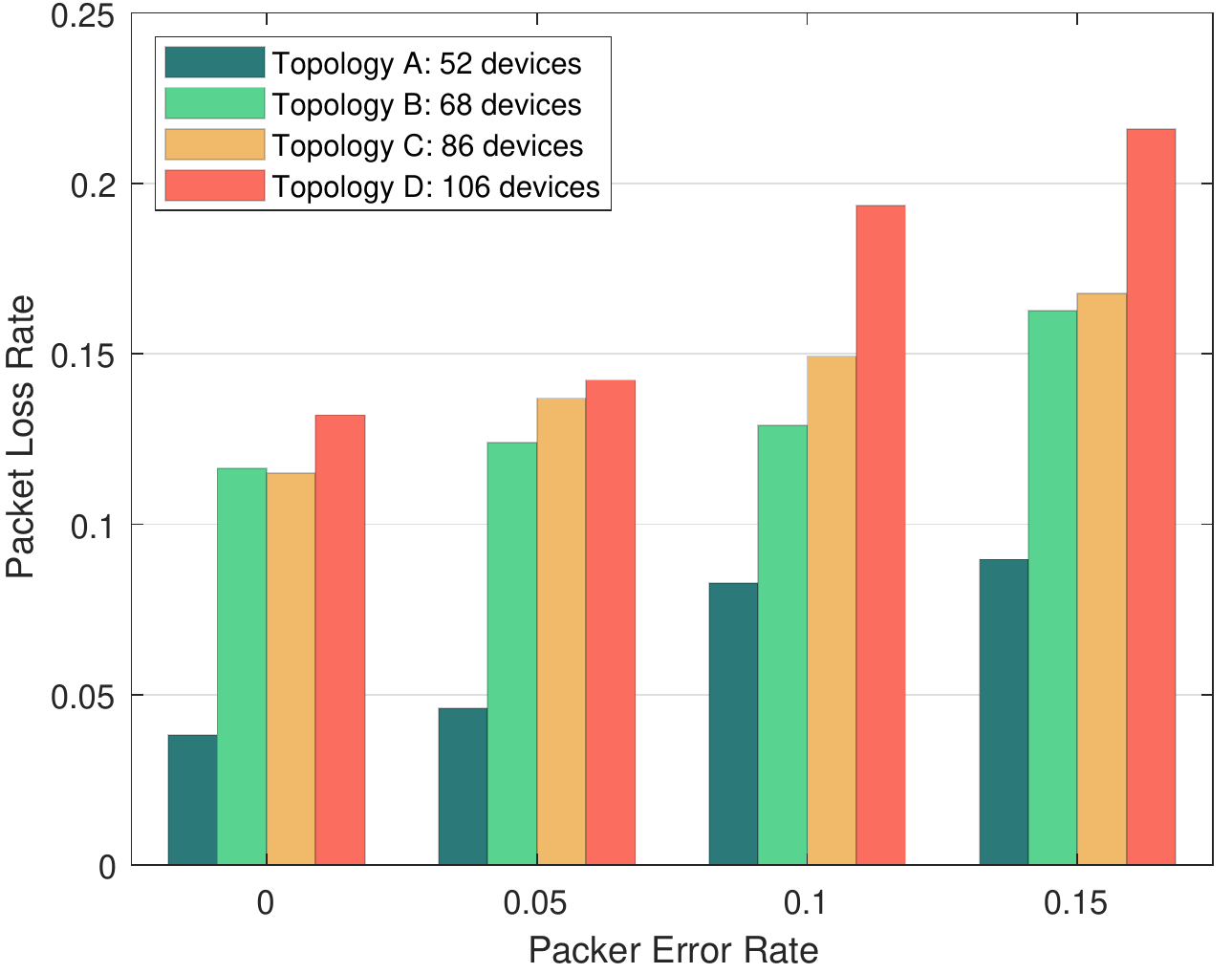} 
        \caption{End-to-end packet loss rate for increasing number of devices, with side traffic and $T_{\mathrm{interPDU}}\sim(3-5)$~ms.}
        \label{MeshR51L}
    \end{minipage}\hfill
    \begin{minipage}{0.49\textwidth}
        \includegraphics[width=0.9\textwidth]{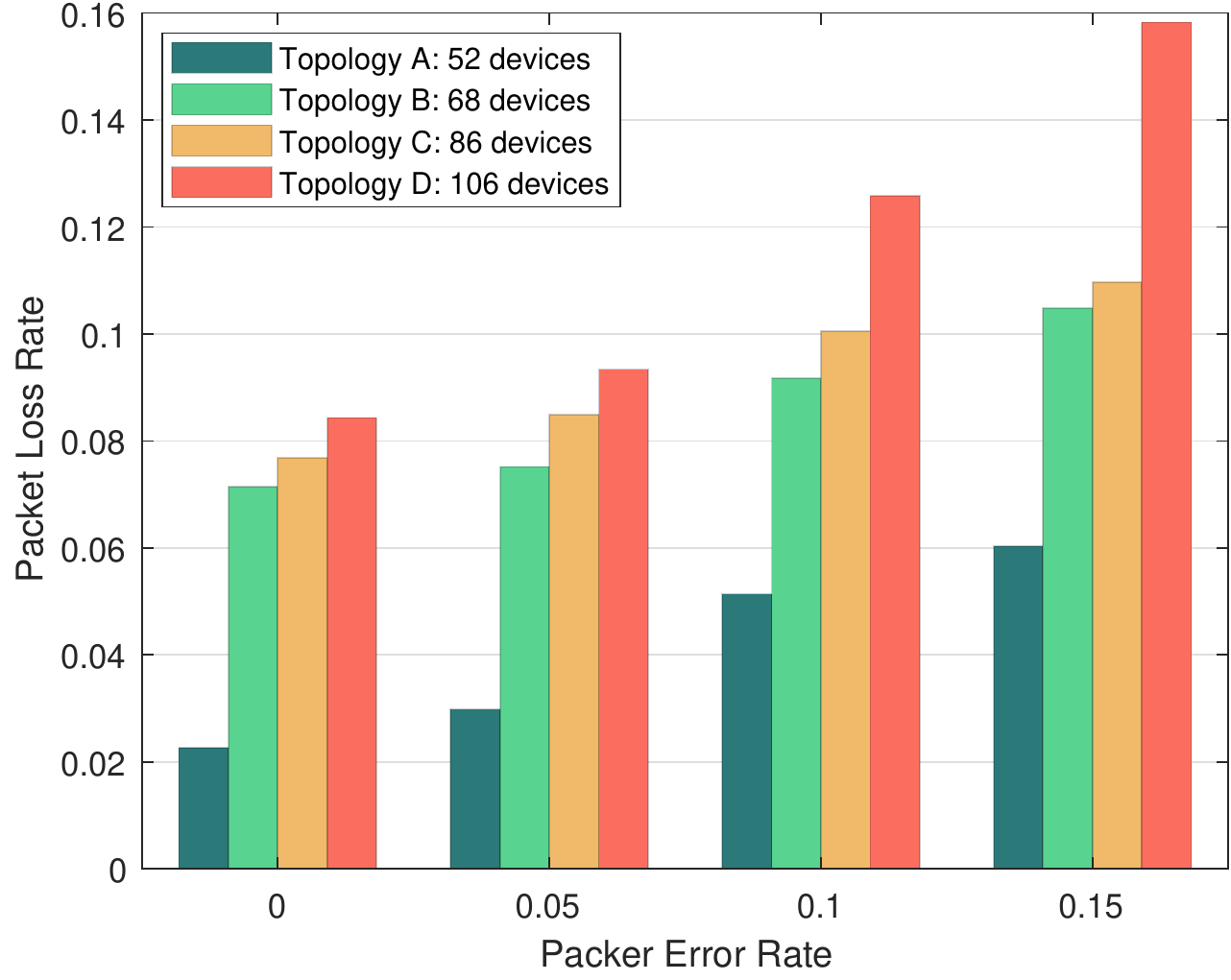} 
        \caption{End-to-end packet loss rate for increasing number of devices, with side traffic and $T_{\mathrm{interPDU}}\sim(0.1-1)$~ms.}
        \label{MeshR51S}
    \end{minipage}
\end{figure*}
 \begin{figure*}[h]
    \centering
    \begin{minipage}{0.49\textwidth}
        \includegraphics[width=0.9\textwidth]{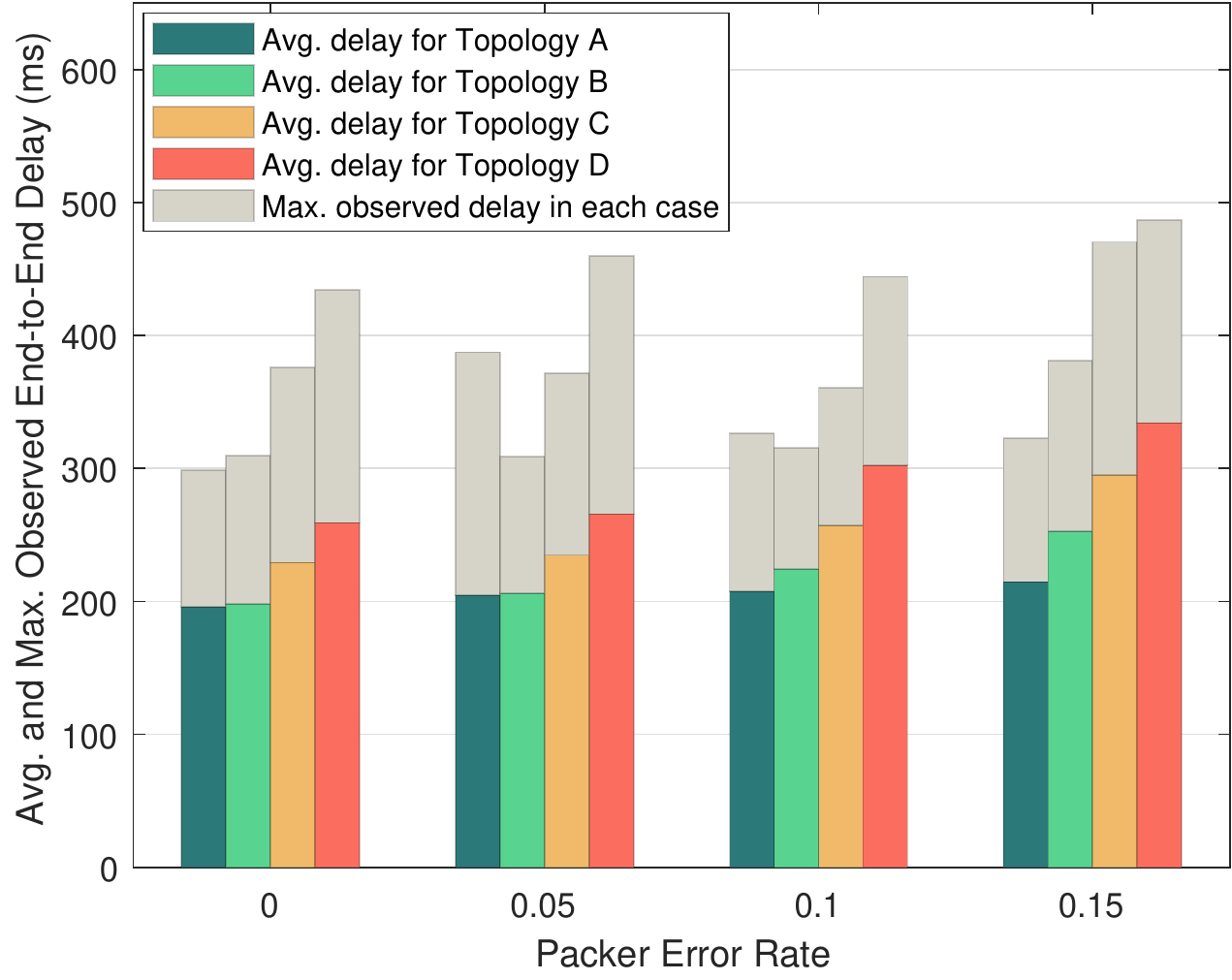} 
        \caption{End-to-end avg. and max. delay for increasing number of devices, with side traffic and $T_{\mathrm{interPDU}}\sim(3-5)$~ms.}
        \label{MeshR52L}
    \end{minipage}\hfill
    \begin{minipage}{0.49\textwidth}
        \includegraphics[width=0.9\textwidth]{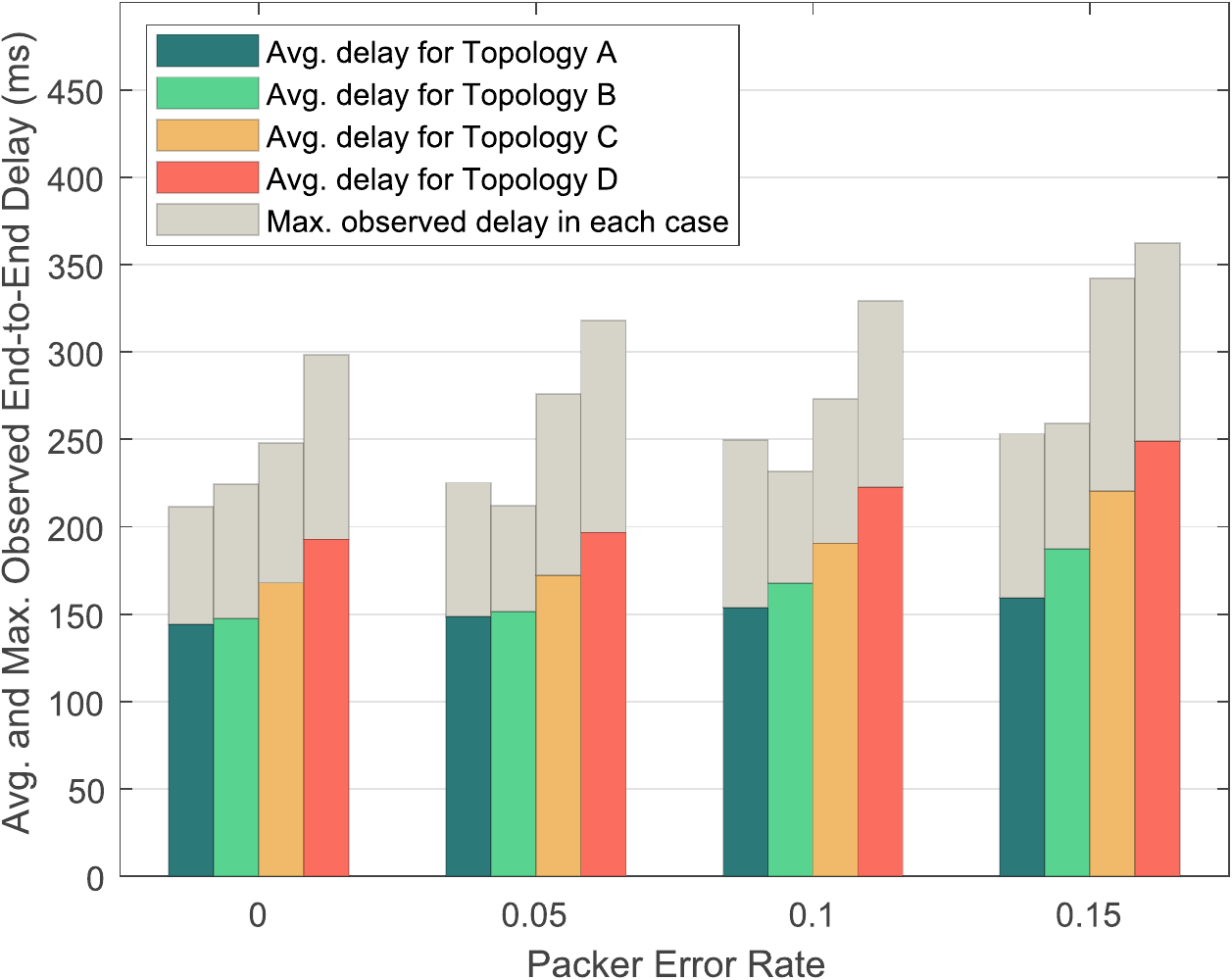} 
        \caption{End-to-end avg. and max. delay for increasing number of devices, with side traffic and $T_{\mathrm{interPDU}}\sim(0.1-1)$~ms.}
        \label{MeshR52S}
    \end{minipage}
\end{figure*}

Due to small PDU size and the randomness introduced before the start and within the span of an \textit{Advertising Event}, the actual collision probability, i.e. two or more devices transmitting on the same channel simultaneously, is remarkably low. However, as previously stated, the choice $T_{\mathrm{interPDU}}$ determines the actual duration of the \textit{Advertising Event}. A device relaying a PDU cannot perform any other task until it has finished the current \textit{Advertising Event}. With a longer time duration of the latter, the probability that other PDUs are not received due to the device being busy with the current PDU, increases. We referred to this as \textit{Congestion Probability}. 

With the inclusion of the side-traffic, the packet loss rate greatly increases in comparison to the previous simulated case. It is shown in Fig. \ref{MeshR51L} that for $T_{\mathrm{interPDU}}$ from 3 to 5~ms, only Topology A shows a packet loss rate below 10\% and that denser topologies perform significantly worse. In Fig. \ref{MeshR51S}, we can see that for all the studied topologies, the use of a $T_{\mathrm{interPDU}}$ from 0.1 to 1~ms results in a decrease in the average e2e packet loss rate of up to 20\%.

A similar trend is observed for the delay performance. As seen in Fig. \ref{MeshR52S}, the use of a shorter $T_{\mathrm{interPDU}}$ results in an average decrease of 25\% in both average and maximum observed e2e delay, when compared to the case of using a lager $T_{\mathrm{interPDU}}$, shown in Fig.~\ref{MeshR52L}. 

\begin{figure*}[h]
    \centering
    \begin{minipage}{0.49\textwidth}
        \includegraphics[width=0.9\textwidth]{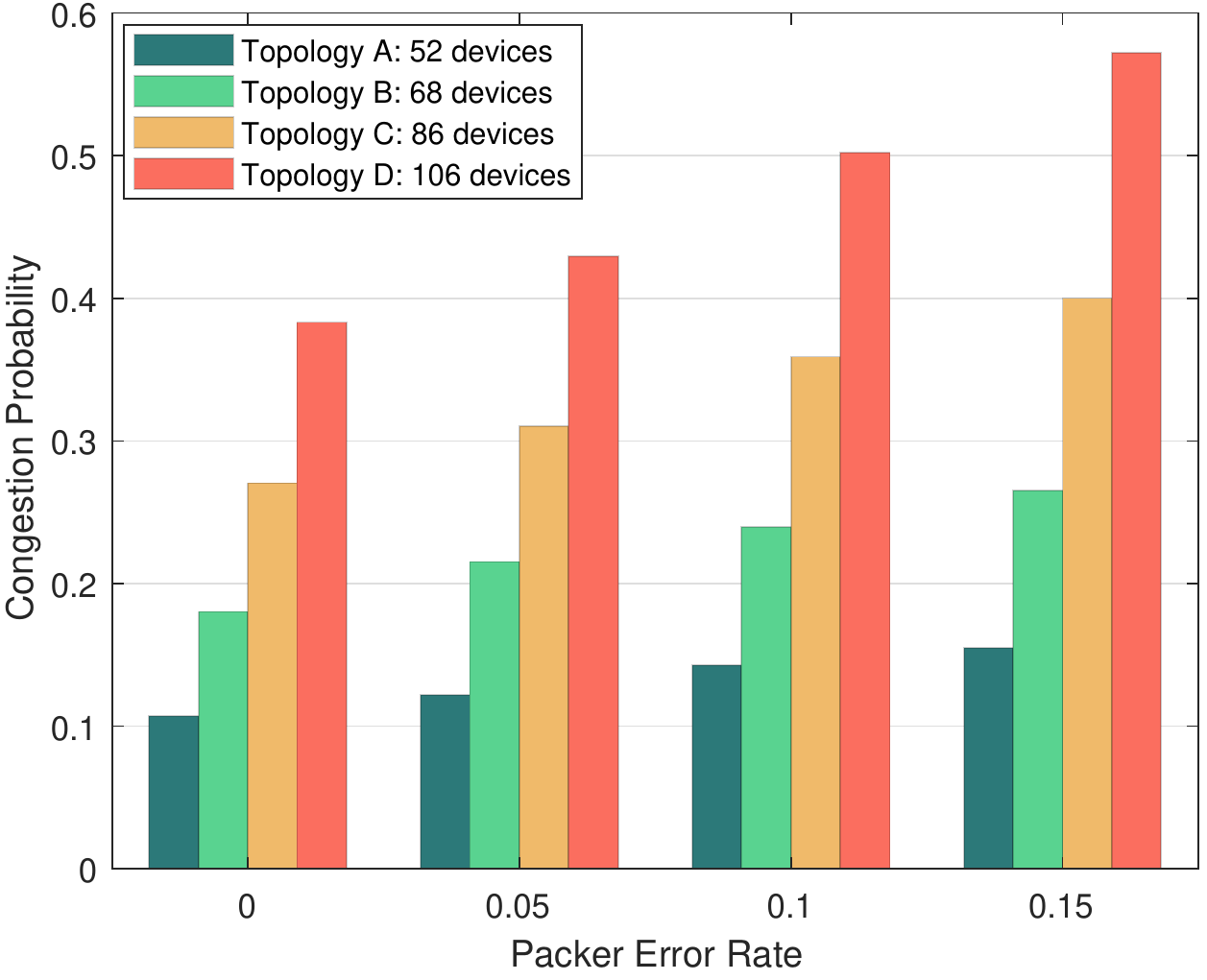} 
        \caption{Congestion Probability for increasing number of devices, with side traffic and $T_{\mathrm{interPDU}}\sim(3-5)$~ms.}
        \label{MeshR53L}
    \end{minipage}\hfill
    \begin{minipage}{0.49\textwidth}
        \includegraphics[width=0.9\textwidth]{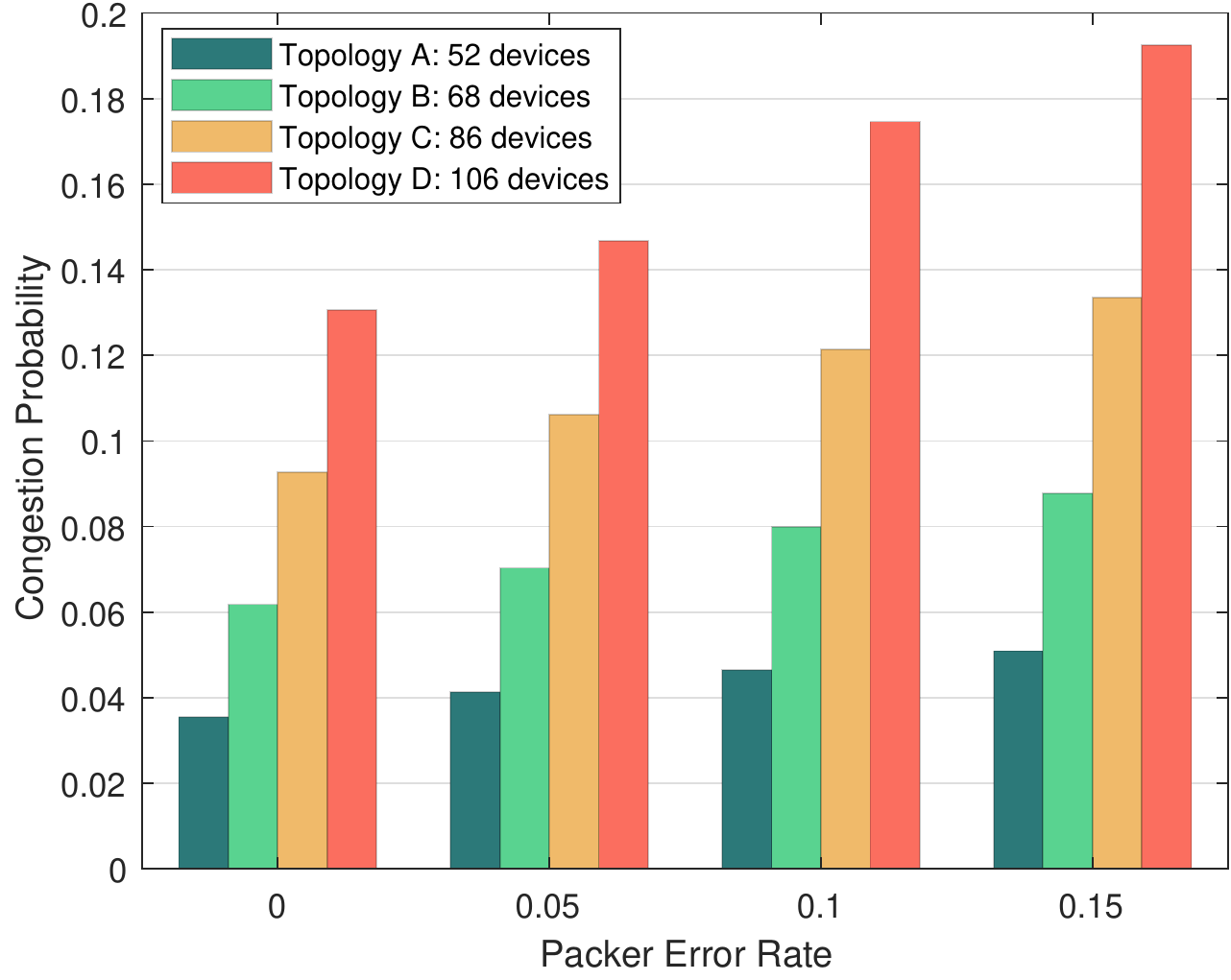} 
        \caption{Congestion Probability for increasing number of devices, with side traffic and $T_{\mathrm{interPDU}}\sim(0.1-1)$~ms.}
        \label{MeshR53S}
    \end{minipage}
\end{figure*}

This behavior is expected since the shorter duration of an \textit{Advertising Event}, which results from the smaller $T_{\mathrm{interPDU}}$, allows the relay devices to handle PDUs faster, thus reducing the congestion and collision probability. We can clearly observe this effect when comparing Fig. \ref{MeshR53S} with Fig. \ref{MeshR53L} where the congestion probability shown in the former is approximately 60\% lower than the one in the latter, for all considered PER conditions. According to these results, the use of shorter $T_{\mathrm{interPDU}}$, when possible, is adviced to improve the reliability and the timeliness of a Bluetooth Mesh network. However, application specific limitations or requirements can restrict the viable ranges for this parameter.  

 \begin{figure*}[h]
    \centering
    \begin{minipage}{0.49\textwidth}
        \includegraphics[width=0.9\textwidth]{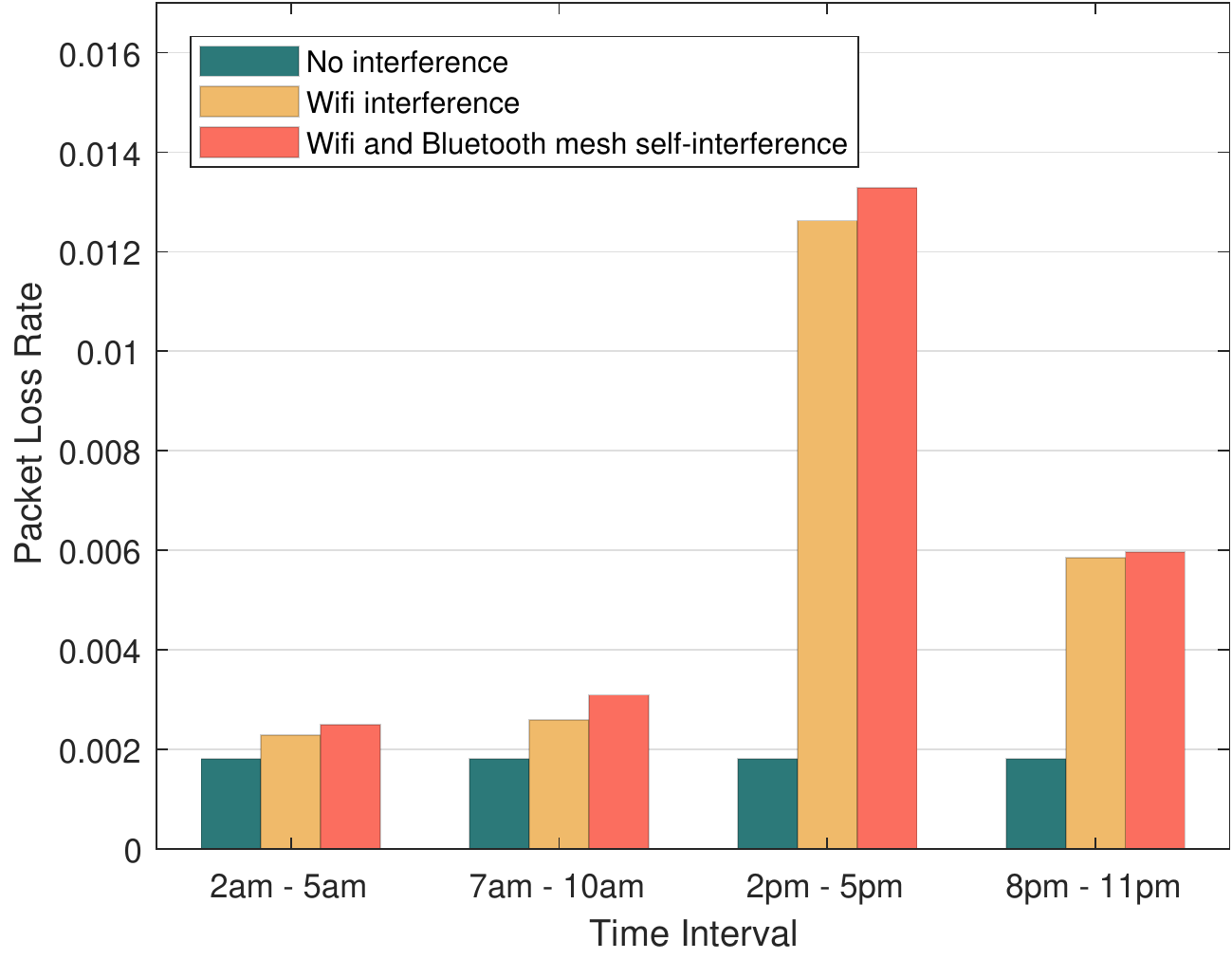} 
        \caption{End-to-end packet loss rate for interference conditions varying throughout the day}
        \label{MeshI1}
    \end{minipage}\hfill
    \begin{minipage}{0.49\textwidth}
        \includegraphics[width=0.88\textwidth]{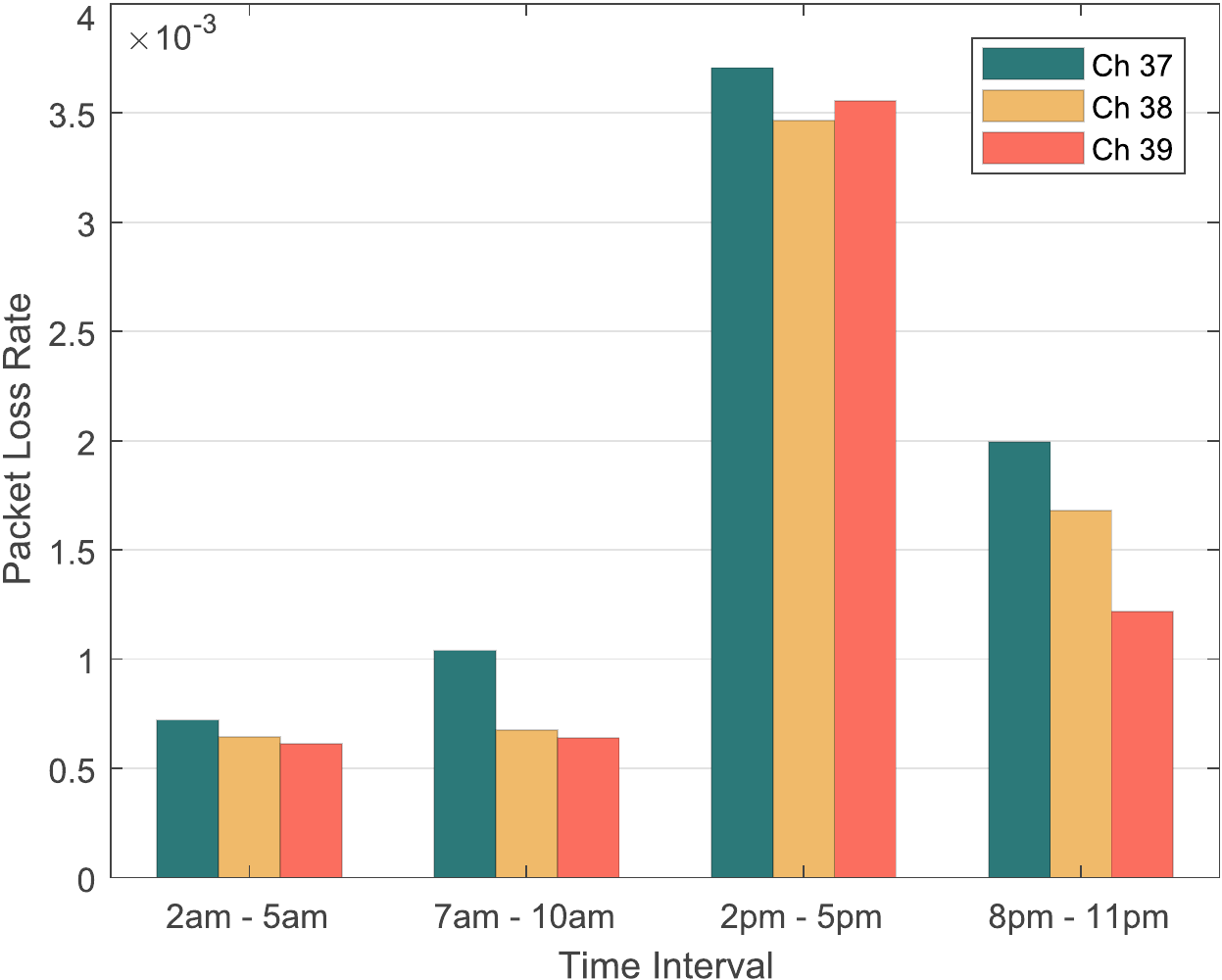} 
        \caption{Packet loss rate per channel for interference conditions varying throughout the day}
        \label{MeshI2}
    \end{minipage}
\end{figure*}

\subsection{Bluetooth Mesh Performance Under WLAN Interference}
\begin{figure*}[!t] 

    \centering
  \subfloat[]{%
       \includegraphics[width=0.9\linewidth]{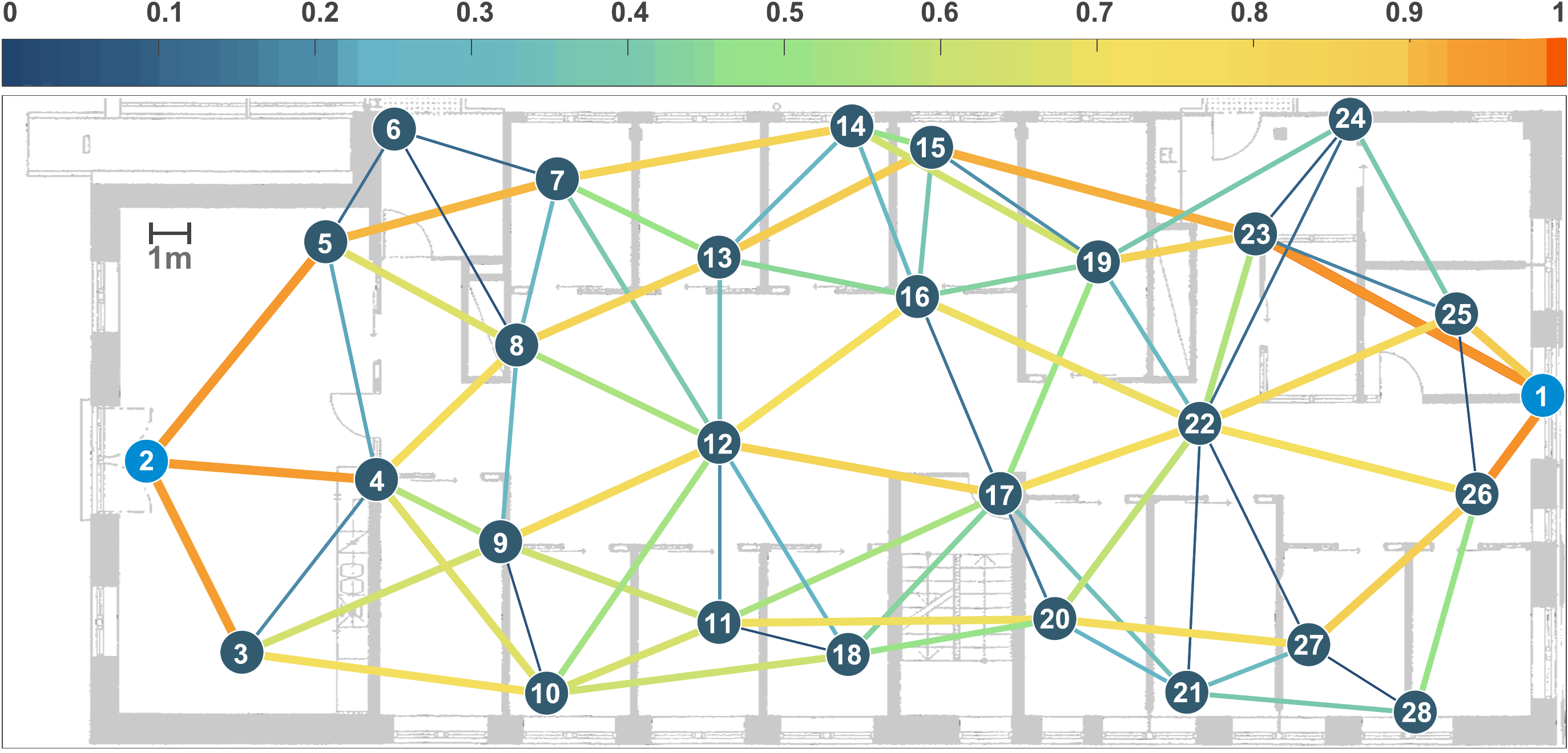}\label{fig:1a}}\hfill
  \subfloat[]{%
        \includegraphics[width=0.9\linewidth]{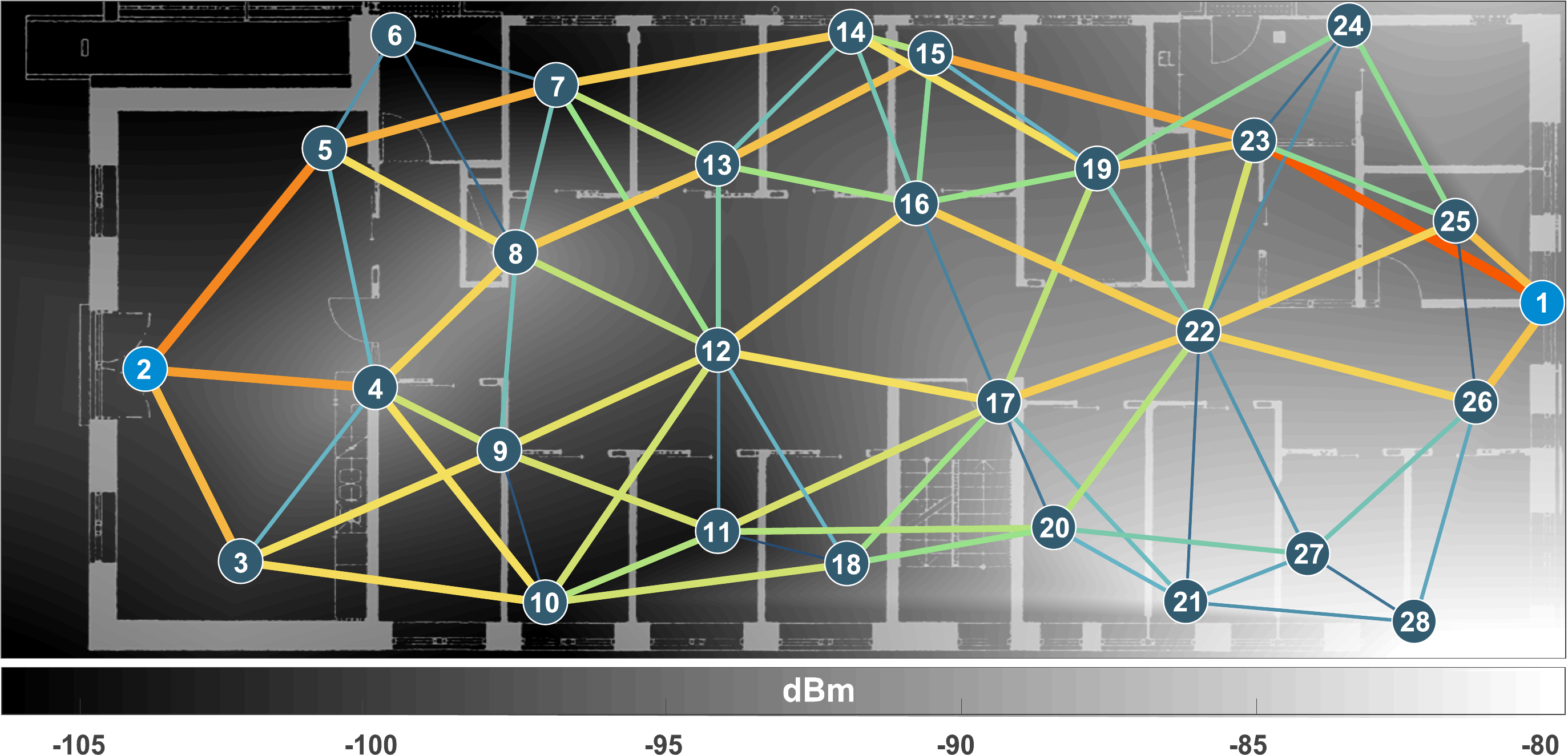}\label{fig:1b}}
  \caption{Point-to-point link successful traffic intensity: (a) Without WLAN interference1. (b) With WLAN interference.}
  \label{fig:MeshInterference}      
\end{figure*}

One of the main limitations of the Bluetooth Mesh protocol is the frequency channel usage. As previously stated, only 3 of the 40 available BLE frequency channels are currently used in Bluetooth Mesh. These three are the ADV channels; channels 37, 38, and 39, used in BLE only for device discovery, connection-establishment procedures, and ADV beacons. These channels are spread across the BLE frequency spectrum to reduce the probability that they are all blocked simultaneously. Channels 37 and 39 are the first and last channels in the frequency band, respectively, while 38 is in the center. Moreover, these channels do not considerably overlap with the frequency channels used in technologies operating in the 2.4 GHz spectrum. In particular, channel 38 is placed between WLAN channels 1 and 6 so to avoid the WLAN interference. 

In this section, we study the behavior of a Bluetooth Mesh network in the presence of interference from WLAN. For this purpose, we designed a mesh network topology to fit a real office space and simulated the data transmission considering the PER corresponding to the existing 
signal-to-interference-plus-noise-ratio (SINR). The SINR is calculated as in \eqref{eqSINR}, where $P_{Rx}$ is the power of the signal received by a scanning device, while $N$, $I_{W}$, and $I_{BM}$ are the noise floor, power of the WLAN interference, and power of the interference from other Bluetooth Mesh devices transmitting on the same channel, respectively. 
\begin{equation}
    \label{eqSINR}
    \textrm{SINR} = \frac{P_{Rx}}{N + I_{W}+I_{BM}}
\end{equation}

To calculate the received power we used the simplified path loss model with log-normal shadowing, as expressed in \eqref{eqRX}. We considered a transmission power $P_{Tx}$ of $0~\textrm{dBm}$, reference distance $d_0=1$ m, and a path-loss exponent $\gamma=3.5$, characteristic of indoors office environments. Taking into consideration the observations reported in \cite{6842666}, we defined the shadowing $\psi_{\textrm{dB}}$ as a Gaussian-distributed random variable with mean zero and standard deviation $\sigma_{\psi_{\textrm{dBm}}}=4~\textrm{dB}$. The wavelength $\lambda$ is dependent on the used channel, and $d$ is the distance in meters between the transmitting and the receiving devices.
\begin{equation}
    \label{eqRX}
    \frac{P_{Rx}}{P_{Tx}}\textrm{dBm}=20\log_{10}\left(\frac{\lambda}{4\pi _0}\right)-10\gamma \log_{10}\left(\frac{d}{d_0}\right)-\psi_{\textrm{dB}}
\end{equation}

The PER for an Additive White Gaussian Noise (AWGN) channel can be estimated with \eqref{eqPER}, in which $n=312$ bits is the PDU size, $\textrm{erfc}(\cdot)$ is the complimentary error function, and $\alpha$ is a constant. In \cite{1095089}, $\alpha=0.68$ was found to be a suitable value for GMSK modulation.
\begin{equation}
    \label{eqPER}
    \textrm{PER} = \left(0.5\cdot \textrm{erfc}\left(\sqrt{\alpha \cdot \textrm{SINR}}\right)\right)^n
\end{equation}

The mesh we simulated consisted of 28 relay nodes spread around the office environment in which the interference data was recorded, as shown in Fig. \ref{fig:MeshInterference}. The interference maps used in the simulation were constructed from experimental data collected through the real-time interference detection and identification framework presented in \cite{bib1}. Such method employs real-time analysis of single interference bursts to enable distributed identification and mapping of external interference from common IoT technologies, according to their power and temporal features. A detailed description of the external interference data is provided in the Appendix.

We considered four different time intervals for which there was a clear wireless traffic load in the environment. For each interval, we simulated the transmission of 10 thousand packets from device 1 to device 2, with a correspondent interference map. In this case, we used the randomization of the timing parameters previously discussed, as well as the transmission of a single replica of the original PDU, with no side-traffic. We considered a noise floor $N=-106$~dBm and a sensitivity level of $-85$~dBm. The rest of the simulation parameters were configured as described in table \ref{SimulationTable}.  

Fig. \ref{MeshI1} shows that, in the presence of external interference, the e2e packet loss rate greatly varies over time. In the specific environment from which the interference data was obtained, the WLAN traffic from 2 pm to 8 pm is at its highest. For these intervals, there is a noticeable increase in the packet loss rate. This is especially evident in the interval from 2 pm to 5 pm, in which the packet loss rate is more than 4 times higher than the values seen for the interval from 7 am to 10 am. In contrast, the performance is not significantly affected by interference from the other Bluetooth Mesh devices in the network, since it is shown that it only results in a small addition to the increase in the packet loss rate produced by the WLAN interference.

In Fig. \ref{MeshI2}, it is shown that the packet loss rate is, in general terms, evenly distributed for the three channels, with no clear particular pattern in the behavior of each frequency channel. However, channel 37 presents a higher packet loss rate than the other two channels. This is explained by the partial frequency overlapping of the channel 37 of BLE with channel 1 of WLAN. Channels 38 and 39 of BLE, in contrast, do not overlap with WLAN channels 6 and 11, respectively. As a result, the performance of channel 38 and 39 is expected to be closely similar.

Finally, Fig. \ref{fig:MeshInterference} shows the successful traffic intensity for each point-to-point link. The results when considering only noise and Bluetooth Mesh self-interference are presented in Fig. \ref{fig:MeshInterference} \subref{fig:1a}, while Fig. \ref{fig:MeshInterference} \subref{fig:1b} shows the effect of the presence of WLAN interference. When comparing the two cases, results indicate that WLAN interference disturbs the point-to-point success rate throughout the network. The highest values reported in the interference map used for this simulation are concentrated close to the bottom-right corner of Fig. \ref{fig:MeshInterference} \subref{fig:1b}. The sub-network 17-20-21-27-28-26-22, which is located in this area, displays the more noticeable changes due to the increased local PER. In particular, in the presence of WLAN interference, the success rate in links 20-27 and 27-26 is significantly lower than in the case of no interference. As a result, the traffic that transits these links in \ref{fig:MeshInterference} \subref{fig:1a}, is spread among the other links, producing a higher traffic intensity in the upper section of the mesh. Links 1-23 and 14-19 are clear examples of this. If the interference conditions of the environment are known, these results could provide valuable information regarding the mesh topology and the optimal placement of the relay nodes.

Overall, these results show that even when using only the three BLE ADV channels, the performance of a Bluetooth Mesh network is still considerably affected by interference from other common wireless technologies, such as WLAN. Since version 5.0 of the Bluetooth Core Specification, the use of secondary ADV channels has been introduced, but it was not available at the time of conducting our simulation and writing this article. Secondary ADV allows for the use of the remaining of the BLE frequency channels for broadcast-oriented communication, which in turn results in greater flexibility for the Bluetooth Mesh protocol. As an example, the results discussed in this section could be improved by using a different set of channels better suited for the conditions of the specific environment.


\section{Conclusion}
In this article, we analyzed the QoS performance of the recently released Bluetooth Mesh protocol. In particular, we identified the most important protocol parameters and their impact on system performance. Our findings show that the main limitation of the protocol is scalability. Bluetooth Mesh is particularly vulnerable to network congestion and increased packet collision probability for densely populated deployments, which is a characteristic of flooding-based networks. The randomization of the timing parameters of the protocol proved to be a solid starting point to counteract network congestion. In addition, we showed that the timing configuration of the \textit{Advertising Events} and the \textit{Scanning Events} present a mutual dependency and that by choosing inadequate conditions, the system does not successfully operate. Moreover, results proved that using smaller values for the $T_{\mathrm{interPDU}}$ for the different PDU transmissions within the \textit{Advertising Events} further reduces the congestion probability, as well as the end-to-end delays and Packet Success Rate.

Finally, we studied the behavior of the protocol when taking into consideration both self and external interference. Our results proved that the performance of Bluetooth Mesh is not notably affected by interference from other Bluetooth Mesh devices in the network. This is especially true with shorter \textit{Advertising Events} and an appropriate configuration for the \textit{Scanning Events}. In contrast, even when using only the three BLE ADV channels, Bluetooth Mesh is still considerably susceptible to interference from other common wireless technologies, such as WLAN. The recent introduction of secondary ADV channels, however, provides a greater degree of flexibility to design the system according to a given environment. This could constitute a key component when trying to enhance the performance of Bluetooth Mesh in the presence of different types of interference. In general terms, channel usage and channel hopping patterns, as well as policies for the spatial distribution of relay nodes, are some of the topics that should be prioritized for future works, as these are critical aspects that could improve the scalability of the protocol.

\appendix[On the Construction of Interference Maps]

The interference map used in this work is constructed utilizing the interference identification framework presented in \cite{bib1}. The identification framework has been implemented in an IEEE 802.15.4-based wireless sensor network (WSN). The WSN has been deployed in an office setting with interference from WLANs composed of three IEEE 802.11b/g/n access points and several static and mobile end-devices.
The interference data collection encompassed the full \SI{2.4}{\giga\hertz} ISM-band and a time-window of approximately \SI{42}{\hour}. The selected time-window allowed us to capture the long-term dynamics of the spectral distribution of interference.
The \textit{a priori} known position of the WSN devices has been exploited to construct a continuous interference power distribution in the space-domain via interpolation. The result of the interpolation process is a multi-dimensional map of interference power in time, frequency, and space domains.
Finally, the obtained map is adapted for usage with the proposed simulator by appropriately selecting snapshots representing different time-windows. In this manner, the simulated BLE mesh network is exposed to various interference conditions according to the fluctuations of the WLANs data-traffic intensity.

\subsection*{Adaptation of Recorded Interference Data}
The employed interference identification solution is based on IEEE 802.15.4-compliant radios and collects information on each of the \SI{16}-channels defined by the IEEE 802.15.4 standard. Since the BLE radio-system employs a different physical layer (PHY), an adaptation of the recorded data is necessary. In particular, we recognize two main aspects: a) the BLE standard defines \SI{40}-channels, with a layout different than the IEEE 802.15.4 allocation \cite{bib3}, and b) the channel width of IEEE 802.15.4 PHY is wider than the BLE standard.

The first factor (i.e., the different channel layout) has been addressed via a surjective channel mapping, which enables to switch between the two channel layouts. In other words, the interference data for each of the \si{40} BLE channels is extrapolated from the observation on the overlapping IEEE 802.15.4 channel.
More elaborated reasoning has to be used to take into account the different channel widths. This aspect is indeed closely related to the bandwidth of the channel filters employed in the different radios. Since the BLE standard defines narrower filter bandwidth than the IEEE 802.15.4 standard \cite{bib2}, the interfering signal power received by an equivalent-gain BLE radio would be inferior. To quantify this variation, we express the total power $P_S$ of a recorded interference burst as
\begin{equation}
P_S = \int_{- \infty}^{\infty} S(f)H(f) \, df,
\label{EQ:avg_power}
\end{equation}
where $S(f)$ is the spectrum of the band-limited interfering signal, and $H(f)$ is frequency response of the employed radio front-end. 

Since $H(f)$ is regulated by the adopted channel filter, let $H_R(f)$ be the frequency response of the hardware used for interference mapping (IEEE 802.15.4) and $H_B(f)$ be the channel filter of the target system (BLE).
It follows that the total power $P_B = \int_{- \infty}^{\infty} S(f)H_B(f) \,df$ observed by a BLE radio would be inferior to the one recorded by the IEEE 802.15.4 system $P_R$, due to the narrower channel filter response.
As we are interested in estimating $P_B$ from the observed $P_R$, we exploit the knowledge that that the bandwidth of the IEEE 802.11 interfering signals (\SI{20}{\mega\hertz} or \SI{40}{\mega\hertz}) is much larger than the  channel filter bandwidths of the observing system, i.e. \SIrange{1}{4}{\mega\hertz}, as described in \cite{bib2}. Under this assumption, we can approximate the spectrum of the wide-band interfering signal as almost flat in the range of non-zero values of both the frequency responses $H_R(f)$ and $H_B(f)$. In other words, we consider $S(f) \approx S(f_c)$, if $\left| f - f_c\right| < {B_{R}}/{2}$, with $f_c$ the central frequency of the observation and $B_{R}$ the bandwidth of the specific channel filter. Then, we can use (\ref{EQ:avg_power}) to rewrite $P_R$ and $P_B$ as
\begin{equation}
P_R = S(f_c) \int_{- \infty}^{\infty} H_R(f) \,df   
\label{EQ:approxA}
\end{equation}
\begin{equation}
P_B = S(f_c) \int_{- \infty}^{\infty} H_B(f) \,df 
\label{EQ:approxB}
\end{equation}
By combining (\ref{EQ:approxA}) and (\ref{EQ:approxB}) it is finally possible to estimate the total equivalent interference signal power affecting a BLE receiver as
\begin{equation}
P_B = P_{R} \cdot \frac{B_{B}}{B_{R}} =  P_{R} R_B,
\label{EQ:PB}
\end{equation}
where $R_B=B_{B}/B_{R}$ is the ratio between the bandwidths of BLE and IEEE 802.15.4 channel filters. The relation in (\ref{EQ:PB}) can be also expressed in logarithmic units as $P_B|_{\text{dBm}} = P_R|_{\text{dBm}} +  10\log_{10}{R_B}$ with $R_B=\SI{6.02}{\decibel}$, which is derived from \cite{bib2} and \cite{bib3}.
%








\bibliographystyle{IEEEtran}
\bibliography{MagBib}

\end{document}